\documentclass[aps,prd,reprint,onecolumn,12pt,tightenlines,raggedbottom,notitlepage,superscriptaddress,nofootinbib,longbibliography]{revtex4-2}
\usepackage{graphicx}
\graphicspath{{figures/}}
\usepackage{amsmath,amssymb,bm,slashed,booktabs}
\usepackage{placeins}
\usepackage{hyperref}
\hypersetup{colorlinks=true,allcolors=blue}
\makeatletter
\def\NAT@sort{\z@}
\makeatother

\begin{document}

\title{Chiral crossover responses in a magnetic field:
Finite-\texorpdfstring{$\mu_5$}{mu5} matching in the NJL model}

\author{Jingyi Chao}%
\email{chaojingyi@jxnu.edu.cn}
\affiliation{School of Physics, Jiangxi Normal University, Nanchang, Jiangxi 330022, People's Republic of China}

\date{November 17, 2026}

\begin{abstract}
We investigate how independent information at finite chiral chemical
potential $\mu_5$ constrains pseudocritical-temperature responses in the
Nambu--Jona-Lasinio model under a magnetic field. Vacuum calibration
alone leaves the coefficient $c_r$ of the additional interaction
$\delta\Omega_r=c_r\mu_5^2(M-m_0)^2$ undetermined. We compare
soft-cutoff (SC) and Pauli--Villars (PV) vacuum regularizations, each
combined with the direct treatment (D) or the magnetic finite-difference
scheme (MFDS). The reference SC and PV treatments exhibit opposite
finite-$\mu_5$ responses already at $eB=0$. Matching the $eB=0$
scalar-susceptibility peak-temperature ratio predicted by unitarized
chiral perturbation theory gives $c_{\rm SC}=-0.07112$ and
$c_{\rm PV}=0.22479$. Holding these coefficients fixed at
$eB=0.1,\,0.2,\,0.4\,\mathrm{GeV}^2$ reduces the four-treatment spread of
the normalized finite-$\mu_5$-induced mass-inflection temperature change
by factors of approximately $7.5$--$15.3$ at $\mu_5=0.15\,\mathrm{GeV}$
and $18.1$--$32.6$ at $\mu_5=0.30\,\mathrm{GeV}$. For the distinct
scalar-susceptibility peak-temperature ratio at $\mu_5=0.15\,\mathrm{GeV}$,
the spread decreases by factors of approximately $74$, $27$, and $14$
at the same magnetic fields, respectively. The reduction in
regularization-dependent finite-$\mu_5$ responses thus persists at the
sampled nonzero magnetic fields without further matching.
\end{abstract}
\maketitle
\enlargethispage{-4pt}

\section{Introduction}\label{sec:introduction}
Strong magnetic fields provide an environment for studying QCD matter under
extreme conditions. They are generated in noncentral relativistic heavy-ion
collisions \cite{Kharzeev:2007jp}, occur in magnetars
\cite{Duncan:1992hi}, and may have been present in the early Universe
\cite{Vachaspati:1991nm}. By modifying the spectra of charged particles and
the dynamics of chiral symmetry breaking, magnetic fields can significantly
affect the chiral crossover in QCD matter
\cite{Andersen:2014xxa}. The resulting changes in chiral
observables offer important tests of low-energy effective descriptions
of QCD matter.

The chiral condensate serves as a central example. Analytical studies
identify dimensional reduction in fermion pairing as a mechanism for
magnetic catalysis \cite{Gusynin:1994xp}. Lattice QCD with
physical quark masses finds an enhanced condensate at low temperature,
but a suppressed condensate and a decreasing pseudocritical temperature
near the crossover as the magnetic field increases
\cite{Bali:2011qj,Bali:2012zg}. This inverse magnetic catalysis (IMC)
reflects competing valence and sea contributions, including the magnetic
response of the gauge ensemble \cite{Bruckmann:2013oba}. These
results set a benchmark for effective quark models and have motivated
lattice-constrained interactions in the Nambu--Jona-Lasinio (NJL)
framework \cite{Farias:2016gmy}.

Axial imbalance introduces another aspect of magnetized QCD matter.
Topological charge-changing transitions can generate an axial charge
asymmetry through the axial anomaly \cite{Kharzeev:2007jp}, while a
finite chiral chemical potential in a magnetic field can give rise to the
chiral magnetic effect \cite{Fukushima:2008xe}. Chiral imbalance has also
been explored as a possible ingredient in descriptions of IMC
\cite{Chao:2013qpa,Liu:2025uey}. Here we focus on the equilibrium response
to a prescribed $\mu_5$. Since axial charge is not exactly conserved, we treat
$\mu_5$ as a quasistatic external parameter; determining its production
and relaxation requires additional dynamical information
\cite{Ruggieri:2020qtq}.

Previous studies have reported qualitatively different dependences of
the pseudocritical temperature on $\mu_5$. Retaining or truncating the
thermal contribution can change the temperature trend \cite{Yu:2015hym},
while medium separation and traditional regularization lead to
contrasting temperature trajectories \cite{Farias:2016let}, including in
calculations beyond leading large
$N_c$ \cite{daSilva:2025koa}. Lattice simulations report an increasing
chiral-transition temperature with $\mu_5$ \cite{Braguta:2015owi}.
Dyson--Schwinger calculations find that the critical-endpoint
temperature increases with $\mu_5$ at finite quark chemical potential
$\mu$ \cite{Xu:2015vna,Shi:2020uyb}, while a nonlocal NJL study finds an increasing
critical temperature with $\mu_5$ for a second-order transition in the chiral
limit \cite{Ruggieri:2016ejz}. These studies address different physical
regimes and transition criteria, while model comparisons also reveal
substantial sensitivity to regularization and the treatment of thermal
contributions. They motivate a closer examination of which aspects of
the finite-$\mu_5$ response are determined by conventional vacuum calibration.

This also raises the question of how the $\mu_5$ dependence is constrained.
At zero magnetic field, a PNJL study introduced a $\mu_5$-dependent
Polyakov-loop potential, fitted to lattice results and combined with
medium separation, to reproduce increasing chiral and deconfinement
temperatures with $\mu_5$ \cite{Azeredo:2024sqc}.
In the external-source formulation of low-energy QCD, symmetry restricts
the allowed operators, while their coefficients must be determined from
suitable observables or correlation functions
\cite{Gasser:1983yg,Gasser:1984gg}. In the NJL model,
vacuum calibration alone cannot determine the couplings of
additional interactions proportional to $\mu_5^2$. For a constant
flavor-singlet axial background, chiral perturbation theory introduces
additional low-energy constants in the pion sector \cite{Espriu:2020dge}.
A related sensitivity
to matching conditions appears in the quark--meson model, where different
finite-$\mu_5$ renormalization conditions can produce opposite
critical-temperature shifts in the chiral limit while preserving the same vacuum
calibration \cite{Ruggieri:2016cbq}. We therefore ask whether
independent information at finite $\mu_5$ can determine an additional
coupling not fixed by vacuum calibration and reduce the treatment
dependence of finite-$\mu_5$ crossover responses at nonzero
magnetic fields.

We address this question by introducing
$\delta\Omega_r=c_r\mu_5^2(M-m_0)^2$ in the effective thermodynamic
potential, with a constant $c_r$ for each vacuum regularization. In the
scalar mean-field potential, this is equivalent to a specific
$\mu_5^2$-dependent effective NJL coupling. Matching the
scalar-susceptibility peak-temperature ratio between
$\mu_5=0.15\,\mathrm{GeV}$ and $\mu_5=0$ at $B=0$, as predicted by
unitarized chiral perturbation theory
\cite{GomezNicola:2023ghi}, gives $c_{\rm SC}=-0.07112$ for
soft-cutoff (SC) regularization and $c_{\rm PV}=0.22479$ for
Pauli--Villars (PV) regularization. For each vacuum regularization, we
compare direct regularization of the Landau-level vacuum contribution
with the magnetic finite-difference scheme (MFDS), in which the
ultraviolet-finite difference between the Landau-level sum and the
corresponding continuum integral at $eB=0$ is added to the separately
regularized
$B=0$ contribution. The same matched $c_r$ is used for both magnetic treatments.

For the original $c_r=0$ treatments, the reference SC and PV
regularizations already exhibit opposite finite-$\mu_5$ temperature
responses at $eB=0$, where D and MFDS coincide within each vacuum
regularization. This contrast persists at nonzero magnetic fields,
while the examined SC vacuum-recalibration families contain both
response signs. Magnetic inverse reconstruction against the normalized
lattice-QCD trend provides an auxiliary diagnostic of model dependence
at $c_r=0$. These results show that vacuum calibration alone does not
establish a common SC/PV finite-$\mu_5$ response. With $c_r$
independently matched for the two vacuum regularizations, the
finite-$\mu_5$-induced mass-inflection pseudocritical-temperature
changes become substantially closer among the four treatments at the
sampled magnetic fields. The scalar-susceptibility peak-temperature
ratio exhibits a similar reduction in spread, by factors of approximately
$74$, $27$, and $14$ at $eB=0.1,\,0.2,\,0.4\,\mathrm{GeV}^2$,
respectively. The two comparisons test the finite-$\mu_5$ constraint
across different crossover criteria and at magnetic fields not used
for matching.

The paper is organized as follows. Section~\ref{sec:model} develops
the NJL model at finite $B$ and $\mu_5$ and specifies the additional
coupling and its independent matching condition.
Section~\ref{sec:mu5B-phenomenology} examines the original
$c_r=0$ temperature response, its vacuum and thermal origins,
and the sensitivity to vacuum calibration, followed by a magnetic
inverse diagnostic. Section~\ref{sec:phase-results} compares the
finite-$\mu_5$-induced mass-inflection temperature changes and
the scalar peak-temperature ratios after matching $c_r$.
Finally, Sec.~\ref{sec:discussion-conclusions} discusses the
scope and limitations of these constraints and possible further
tests. Supporting derivations and matching-input sensitivity checks
appear in the appendices.

\section{Model and computational framework}\label{sec:model}
We formulate the equilibrium problem in magnetic and axial backgrounds
to compare responses in models calibrated to the same vacuum observables.
We distinguish the vacuum regularization at $B=0$ from its
extension to finite magnetic fields.

\subsection{Quark propagator and mean-field gap equation}\label{sec:njl-background}
Our starting point is the two-flavor local NJL Lagrangian
\cite{Klevansky:1992qe,Buballa:2003qv},
\begin{equation}
\mathcal L={}\bar\psi(i\slashed D+\mu_5\gamma^0\gamma^5-m_0)\psi
+G[(\bar\psi\psi)^2+(\bar\psi i\gamma^5\boldsymbol\tau\psi)^2],
\label{eq:NJL-Lagrangian}
\end{equation}
where $D_\mu=\partial_\mu+iq_fA_\mu$ acts diagonally in flavor space,
with $q_u=2e/3$ and $q_d=-e/3$. We take $N_c=3$ and a uniform magnetic
field along $z$. The common current mass is $m_0$. In the Hartree
approximation the constituent mass is
$M=m_0-2G\sum_f\langle\bar\psi_f\psi_f\rangle$. We consider the homogeneous scalar channel; competing spatially
modulated or additional condensates lie outside this calculation.

\label{sec:propagator-main}
The Landau-level construction in Appendix~\ref{app:propagator}
determines the quark excitation energies and their contributions
to the chiral condensate.
For nonzero magnetic field, we write $\mathcal B_f=|q_fB|$ and
$s_f=\operatorname{sgn}(q_fB)$. Factoring out the Schwinger phase~\cite{Miransky:2015ava} gives the
translation-invariant momentum-space propagator
\begin{align}
\widetilde S_f(\omega,p_z,\bm p_\perp)
&=ie^{-\alpha_f/2}\sum_{n=0}^{\infty}(-1)^n
\frac{\mathcal K_{n,+}^f+\mathcal K_{n,-}^f}{U_n^f},
\label{eq:propagator-master}\\
\mathcal K_{n,+}^f&=\big[-2L_{n-1}(\alpha_f)\Lambda_L^f+4L_{n-1}^{1}(\alpha_f)\Lambda_T^f\big]P_+^f,\notag\\
\mathcal K_{n,-}^f&=\big[2L_n(\alpha_f)\Lambda_L^f+4L_{n-1}^{1}(\alpha_f)\Lambda_T^f\big]P_-^f.
\label{eq:Kn-plus}
\end{align}
Here $\omega$ is the frequency, $\alpha_f=2p_\perp^2/\mathcal B_f$,
and $P_\pm^f=(1\pm is_f\gamma^1\gamma^2)/2$ are the spin projectors.
The functions $L_n$ and $L_n^1$ are Laguerre and generalized Laguerre
polynomials, respectively, with negative-index polynomials set to zero.
The denominator is $U_n^f=(X_n^f)^2-4\mu_5^2(\omega^2-M^2)$,
where $X_n^f=\omega^2-p_z^2+\mu_5^2-M^2-2n\mathcal B_f$.
The numerator factors contain six longitudinal and three transverse
Dirac structures:
\begin{align}
\Lambda_L^f={}&M(X_n^f-2\mu_5^2)
+\omega(X_n^f-2\mu_5^2)\gamma^0-p_zX_n^f\gamma^3\notag\\
&-\mu_5\big[X_n^f-2(\omega^2-M^2)\big]\gamma^0\gamma^5-2\mu_5p_z\omega\gamma^3\gamma^5
-2\mu_5p_zM\gamma^0\gamma^3\gamma^5,
\label{eq:LambdaL-corrected}\\
\Lambda_T^f={}&X_n^f(\gamma_\perp\!\cdot p_\perp)
+2\mu_5\omega(\gamma_\perp\!\cdot p_\perp)\gamma^5-2\mu_5M(\gamma_\perp\!\cdot p_\perp)\gamma^0\gamma^5.
\label{eq:LambdaT-corrected}
\end{align}
Here $\gamma_\perp\!\cdot p_\perp\equiv\gamma^1p_x+\gamma^2p_y$.
The variable $\bm p_\perp$ is conjugate to the transverse separation
after phase factorization.  It describes the orbital kernel in
Eq.~\eqref{eq:propagator-master}; it is not an additional transverse
energy quantum number alongside $n$.
For $n\geq1$, the positive-frequency zeros of $U_n^f$ give two
branches $a=\pm1$. At $n=0$, the lowest Landau level (LLL) has a single allowed
spin projection. The reduced propagator is
\begin{equation}
\widetilde S_{f,\mathrm{LLL}}
=2ie^{-\alpha_f/2}
\frac{\big[\omega\gamma^0-(p_z+s_f\mu_5)\gamma^3+M\big]P_-^f}
{\omega^2-(E_0^f)^2+i0^+}.
\label{eq:LLL-propagator-reduced}
\end{equation}
Here $E_0^f$ is the positive LLL energy defined in
Eq.~\eqref{eq:LLL-energy}.
A common numerator factor cancels one factor of
$U_0^f$, leaving its physical pole pair,
as shown in Appendix~\ref{appA:lll-reduction}.
With $S_{n,f}=\sqrt{p_z^2+2n\mathcal B_f}$, the resulting positive
pole energies are
\begin{align}
E_0^f(p_z)&=\sqrt{M^2+(p_z+s_f\mu_5)^2},\label{eq:LLL-energy}\\
E_{n,a}^f(p_z)&=\sqrt{M^2+(S_{n,f}+a\mu_5)^2},\qquad n\ge1.
\label{eq:hll-energies}
\end{align}
The LLL momentum is signed and integrated over the full real axis.
These energies agree with the Landau spectrum of Ref.~\cite{Sheng:2017lfu}.
The Feynman pole prescription is specified in
Appendix~\ref{appA:lll-reduction}. Appendix~\ref{app:propagator}
derives the propagator from the orbital wave functions, spin
projectors, ordered matrix inversion, and Schwinger phase
\cite{Gorbar:2008hu}.

\label{sec:gap-main}\label{sec:thermodynamic-closure}
Writing $\Phi^{(s)}=\sum_f\langle\bar\psi_f\psi_f\rangle^{(s)}
=-MJ_1^{(s)}$ for the total chiral condensate, the Hartree mass
relation gives the standard NJL mean-field gap equation,
\begin{equation}
\frac{M-m_0}{2G}-MJ_1^{(s)}(M;T,B,\mu_5)=0.
\label{eq:gap-equation-main}
\end{equation}
Here the superscript $(s)$ labels the four complete treatments SC-D, SC-F,
PV-D, and PV-F defined in Sec.~\ref{sec:regularization-new}.
We solve the gap equation directly at each $T$, $B$, and $\mu_5$ to
obtain the candidate constituent masses. Equation~\eqref{eq:gap-equation-main}
applies before introducing the additional coupling $c_r$.

The function $J_1^{(s)}$ is obtained from the coincident quark
propagator. Only the scalar term and the term proportional to
$\gamma^0\gamma^3\gamma^5$ contribute to the spin-projected Dirac trace;
the transverse contribution has zero scalar trace. After the orbital
integration and decomposition into physical energy poles, each allowed
branch, including the LLL, has scalar numerator $2M$
[Eqs.~\eqref{eq:appA-numerator-traces}--\eqref{eq:appA-integrated-scalar-trace}].
The fermionic Matsubara sum supplies $[1-2n_F(E)]/(2E)$ for each pole,
where $n_F(E)=(e^{E/T}+1)^{-1}$. Before applying the vacuum
regularization, the resulting function is
\begin{equation}
J_1={}N_c\sum_f\frac{\mathcal B_f}{2\pi}
\int_{-\infty}^{\infty}\frac{dp_z}{2\pi}
\bigg[\frac{1-2n_F(E_0^f)}{E_0^f}
+\sum_{n=1}^{\infty}\sum_{a=\pm1}
\frac{1-2n_F(E_{n,a}^f)}{E_{n,a}^f}\bigg].
\label{eq:J1-final}
\end{equation}
Only its vacuum contribution is regularized; the thermal contribution
is retained over the full phase space. Section~\ref{sec:regularization-new}
specifies the four treatments of $J_1$.

The quark determinant provides a consistency check: for any physical
energy branch, differentiating its contribution to the thermodynamic
potential gives the same mass weight,
\begin{equation}
\partial_M\!\left[E+2T\ln(1+e^{-E/T})\right]
=\frac{M}{E}\,[1-2n_F(E)].
\label{eq:detail-thermal-mass-weight}
\end{equation}
The derivative holds $T,B,\mu_5$ fixed. Thus the propagator trace and
thermodynamic derivative yield the same gap equation.

When the gap equation yields multiple candidate masses, we compare
their thermodynamic potentials to identify the favored branch. For
this comparison we denote the left-hand side of
Eq.~\eqref{eq:gap-equation-main} by
\begin{equation}
R^{(s)}(M;T,B,\mu_5)
\equiv\frac{M-m_0}{2G}-MJ_1^{(s)}.
\label{eq:gap-residual-main}
\end{equation}
Since $R^{(s)}=\partial_M\Omega^{(s)}$, integrating $R^{(s)}$
over $M$ at fixed $T$, $B$, and $\mu_5$ yields the relative
thermodynamic potential \cite{Moreira:2010bx},
\begin{equation}
\Omega^{(s)}(M;T,B,\mu_5)-\Omega^{(s)}(M_{\rm ref};T,B,\mu_5)
=\int_{M_{\rm ref}}^M R^{(s)}(M';T,B,\mu_5)\,dM'.
\label{eq:relative-Omega-new}
\end{equation}
We evaluate this potential at the mass solutions already obtained
from the gap equation and select the lowest-potential solution among
the resolved candidates, with the search boundaries checked separately.

At finite $\mu_5$, the external axial source permits additional
source-dependent interactions in the NJL model that are not fixed by
vacuum calibration. In chiral perturbation theory, the external-source
construction introduces an $\mathcal O(p^4)$ interaction proportional to
$\mu_5^2\operatorname{tr}(\chi^\dagger U+U^\dagger\chi)$, with an
independent low-energy constant \cite{Espriu:2020dge}. This framework was
subsequently used in unitarized chiral perturbation theory to predict
the $\mu_5$ dependence of the scalar-susceptibility peak temperature
\cite{GomezNicola:2023ghi}. A structurally related term
proportional to $\mu_5^2\sigma^2$ also appears in the renormalization
of the quark--meson effective potential \cite{Ruggieri:2016cbq}.
Motivated by these results, we introduce a single phenomenological
correction to the scalar--pseudoscalar invariant,
\begin{equation}
\delta\Omega_r=c_r\mu_5^2\rho^2,\qquad
\rho^2=\sigma^2+\boldsymbol\pi^2.
\label{eq:matching-source-example}
\end{equation}
Here $r\in\{\mathrm{SC},\mathrm{PV}\}$ labels the vacuum
regularization, and the same coefficient $c_r$ is used for its D and MFDS
treatments. The auxiliary fields are normalized so that $\sigma=M-m_0$ and
$\boldsymbol\pi=0$ in the scalar mean-field sector, giving
$\delta\Omega_r=c_r\mu_5^2(M-m_0)^2$. In this normalization,
$c_r$ is dimensionless. This additional term is phenomenological rather
than derived from the original NJL quark loop. It vanishes at
$\mu_5=0$, leaving vacuum calibration unchanged; its coefficient is
determined using finite-$\mu_5$ information in
Sec.~\ref{sec:source-input}.

\subsection{Two vacuum regularizations and magnetic-field treatments}
\label{sec:regularization-new}
We compare two vacuum regularizations, soft-cutoff (SC) and
Pauli--Villars (PV), each with two treatments of the magnetic-field
contribution. The direct treatment (D) regularizes the Landau-level
vacuum contribution, while the magnetic finite-difference scheme
(MFDS, denoted F) separates the finite magnetic contribution from the
regularized $B=0$ baseline. SC weights the unshifted momentum, whereas
PV combines contributions with regulator masses:
\begin{align}
f_\Lambda(k^2)&=[1+(k^2/\Lambda^2)^5]^{-1},\qquad\text{(SC)}
\label{eq:soft-window-new}\\
M_i^2&=M^2+a_i\Lambda^2,\qquad
(C_i)=(1,-2,1),\qquad (a_i)=(0,1,2),\qquad\text{(PV)}.
\label{eq:PV-data-new}
\end{align}
The unshifted argument is $k^2=p_z^2+2n\mathcal B_f$.
In the direct treatment, these vacuum regularizations act on
each branch of Eq.~\eqref{eq:J1-final} as
\begin{equation}
\frac{1-2n_F(E)}{E}\longrightarrow
\begin{cases}
[f_\Lambda(k^2)-2n_F(E)]/E,&\mathrm{SC}_{\mathrm D},\\
\displaystyle\sum_i C_i/E_i-2n_F(E)/E,&\mathrm{PV}_{\mathrm D}.
\end{cases}
\label{eq:direct-branch-prescriptions}
\end{equation}
Here $E_i$ differs from $E$ only through the replacement
$M^2\to M_i^2$. Since $\partial_M E_i=M/E_i$ at fixed $\Lambda$,
$\partial_M\sum_iC_iE_i=M\sum_i\frac{C_i}{E_i}$.
The PV vacuum determinant therefore retains the physical mass in
the scalar numerator, consistent with the trace and the gap equation
in Eq.~\eqref{eq:gap-equation-main}. In both cases the thermal term contains only the
physical energy and is integrated over the complete phase space,
without a model cutoff or thermal PV replicas
\cite{Pauli:1949zm,Avancini:2020xqe}.

In MFDS (F), the same ultraviolet-finite difference between
the Landau-level sum and the $B=0$ continuum integral is added
to each separately regularized $B=0$ baseline:
\begin{equation}
J_{1,r_{\mathrm F}}(M;T,B,\mu_5)
=J_{1,r}(M;T,0,\mu_5)
+\Delta J_{1,\mathrm{mag}}(M;T,B,\mu_5).
\label{eq:2x2-finite-difference}
\end{equation}
The finite difference includes both vacuum and complete thermal
contributions and is the same for SC-F and PV-F at identical
$M,T,B,\mu_5$. Their equilibrium states can nevertheless differ
because their $B=0$ baselines differ. Related magnetic subtraction
methods are discussed in
Refs.~\cite{Allen:2015paa,Avancini:2018svs,Avancini:2019wed}, whereas MFDS
here denotes the finite-$\mu_5$ construction specified by
Eq.~\eqref{eq:2x2-finite-difference}, with the $B=0$ continuum
contribution evaluated at the same $\mu_5$ and the stated SC/PV
baselines. Appendix~\ref{app:magnetic-difference}
derives the finite difference and its measure. The SC-F and PV-F
comparison therefore isolates differences that remain when the
finite magnetic contribution is shared.

The reference SC and PV parameter sets are separately calibrated at
$T=B=\mu_5=0$ to $m_\pi=138\,\mathrm{MeV}$, $f_\pi=92.4\,\mathrm{MeV}$,
and the single-flavor condensate
$\langle\bar uu\rangle^{1/3}=-250\,\mathrm{MeV}$.
Table~\ref{tab:vacuum-parameters} lists the fitted parameters.
The constituent-quark mass $M_0$ is an output, so equal vacuum inputs
do not require identical masses in the two regularizations.
Both magnetic-field treatments use the corresponding vacuum parameters
without refitting. The calibration formulas and normalization
conventions are given in Appendix~\ref{app:detail-vacuum-inputs}.

\begin{table}[t]
\caption{Reference vacuum calibrations at $T=B=\mu_5=0$. The two regularizations
reproduce the same pion mass, decay constant, and single-flavor
condensate. The constituent mass $M_0$ is an output. Each row supplies
the parameters for both the D and F magnetic-field treatments.}
\label{tab:vacuum-parameters}
\begin{ruledtabular}
\begin{tabular}{lcccc}
Baseline & $m_0$ (MeV) & $G$ (GeV$^{-2}$) & $\Lambda$ (MeV) & $M_0$ (MeV)\\
\hline
SC & 5.2217 & 5.2629 & 618.96 & 334.15\\
PV & 5.2218 & 3.7078 & 863.32 & 236.96
\end{tabular}
\end{ruledtabular}
\end{table}

\subsection{Crossover observables and finite-\texorpdfstring{$\mu_5$}{mu5} matching}
\label{sec:crossover-observables}\label{sec:source-input}
Writing $x=\mu_5^2$, the equilibrium mass $M_*(T,x)$ at fixed magnetic
field and model parameters defines two crossover criteria.
We first use its inflection \cite{Zhang:2016qrl}:
\begin{align}
T_{\rm pc}^{(1)}(eB,\mu_5)&=\underset{T}{\operatorname{arg\,max}}\,|M_T|,
\label{eq:Tpc-D1}\\
K_5(eB,\mu_5)&=\left.\frac{\partial T_{\rm pc}^{(1)}}{\partial(\mu_5^2)}\right|_{eB},\qquad
\kappa_5(eB)=-T_0(eB)K_5(eB,0).
\label{eq:K5-surface}
\end{align}
Here $M_T=(\partial M_*/\partial T)_{x,eB}$ and
$T_0(eB)=T_{\rm pc}^{(1)}(eB,0)$. Thus
$T_{\rm pc}/T_0=1-\kappa_5(\mu_5/T_0)^2+\mathcal O(\mu_5^4)$:
a positive $K_5(eB,0)$, or equivalently a negative $\kappa_5(eB)$,
indicates an initial increase in the pseudocritical temperature
with $\mu_5^2$. Unless specified otherwise, $T_{\rm pc}$
denotes criterion (1). At the selected nondegenerate interior maximum
of $|M_T|$ on a smooth decreasing branch, $M_{TT}=0$ and $M_{TTT}>0$.

The second criterion locates the mass-susceptibility peak.
For related chiral-susceptibility analyses in a magnetic field, see
Ref.~\cite{Das:2019crc}.
For the original baselines, at fixed $G$ and $\Lambda$,
\begin{equation}
\chi_m=\left(\frac{\partial M_*}{\partial m_0}\right)_{T,B,\mu_5}
=\frac{1}{2G R_M},\qquad
T_{\rm pc}^{(2)}=\underset T{\operatorname{arg\,max}}\,\chi_m.
\label{eq:chi-m-new}
\end{equation}
For a specified treatment, we abbreviate the gap residual
$R^{(s)}$ of Sec.~\ref{sec:njl-background} as $R$.
Here $R_M=(\partial R/\partial M)_{T,x,eB}$ is evaluated at $M=M_*$.
The scalar susceptibility is defined by
$\chi_{\rm scal}=-(\partial\Phi/\partial m_0)_{T,B,\mu_5}
=-d^2\Omega_*/dm_0^2$, where $\Phi$ is the flavor-summed quark
condensate and $\Omega_*$ is the potential evaluated on the
equilibrium branch. For the original $c_r=0$ treatments,
$\chi_{\rm scal}=(\chi_m-1)/(2G)$, so the two susceptibilities
have the same peak temperature. This temperature can differ from the
mass inflection because $M_T=-R_T/R_M$ contains the
temperature-dependent numerator $R_T=(\partial R/\partial T)_{M,x,eB}$
along equilibrium.

We use criterion (2) to determine $c_r$ from the
scalar-susceptibility peak-temperature ratio
\begin{equation}
\mathcal R_\chi^{(s)}(eB)=
\frac{T_{\rm pc}^{(2),s}(eB,0.15\,\mathrm{GeV})}
     {T_{\rm pc}^{(2),s}(eB,0)}.
\label{eq:Rchi-proof}
\end{equation}
For each treatment $s$, criterion (2) is used at both values of
$\mu_5$. This finite-$\mu_5$ ratio is distinct
from the local mass-inflection derivative $K_5$.

The input at $B=0$ is taken from the $f_0(500)$ saturation
calculation in unitarized chiral perturbation theory
\cite{GomezNicola:2023ghi}. That calculation uses
low-energy constants fitted to lattice data; here we use its
effective-theory prediction for the scalar peak-temperature ratio,
rather than a direct lattice measurement at finite $\mu_5$. We adopt the common
$B=0$ input $y_0=1.0057$ for the ratio in Eq.~\eqref{eq:Rchi-proof}.
The values $1.0035$ and $1.0079$ define lower and upper nonstatistical
input scenarios. The matching uses the same magnitude of $\mu_5$ in
the NJL model and the ChPT input. The figure extraction and vacuum-input
differences are detailed in Appendix~\ref{app:finite-source-input}.

Within the scalar mean-field potential, the quadratic term can be
rewritten exactly in terms of
\begin{equation}
G_{\mathrm{eff},r}(\mu_5)
=\frac{G_r}{1+4G_rc_r\mu_5^2},\qquad
G_{\mathrm{eff},r}(0)=G_r.
\label{eq:effective-Gmu5}
\end{equation}
The coefficient $c_r$ is constant, so the $\mu_5^2$ dependence
of $G_{\mathrm{eff},r}$ is fully specified without further fitting.
This equality is exact for the scalar mean-field potential and does
not imply an identical modification of other interaction channels.
The phenomenological contribution modifies the gap equation to
\begin{equation}
\frac{M-m_0}{2G_r}-MJ_1^{(s)}
+2c_r x(M-m_0)=0,
\label{eq:gap-equation-matched}
\end{equation}
The added potential contributes
$2c_r\mu_5^2(M-m_0)$ to the gap equation, while the quark-loop
function $J_1^{(s)}$ retains its original definition.
The constituent mass is still determined by solving this modified
gap equation. Every $\mu_5=0$ equilibrium mass trajectory remains
unchanged, while derivatives with respect to $x$ may change even
at $x=0$. Related source-dependent couplings are discussed in
Refs.~\cite{Ruggieri:2016cbq,Espriu:2020dge}.

The scalar susceptibility is evaluated from this same extended
potential, holding $G,\Lambda,c_r$ fixed under current-mass
differentiation.
Writing $A=(2G)^{-1}+2c_rx$, we obtain
\begin{equation}
\chi_m^{(c)}=\frac{A}{R_M^{(c)}},\qquad
\chi_{\rm scal}^{(c)}=-\frac{d^2\Omega_*^{(c)}}{dm_0^2}
=\frac{A^2}{R_M^{(c)}}-A.
\label{eq:matched-chi-m-app}
\end{equation}
Appendix~\ref{app:matching-operators} derives these expressions.
In the sampled domain, $A$ is positive and independent of $T$;
hence $\chi_{\rm scal}^{(c)}=A(\chi_m^{(c)}-1)$ and
$\chi_m^{(c)}$ peak at the same temperature.
The matching calculation evaluates $\chi_{\rm scal}^{(c)}$ directly
and locates its peak by bounded maximization along the selected
equilibrium branch. Equation~\eqref{eq:detail-susceptibility-locator}
characterizes this common peak analytically; the two susceptibilities
do not supply separate matching conditions.

For each $r=\mathrm{SC,PV}$, we fix the vacuum parameters in
Table~\ref{tab:vacuum-parameters} and solve the extended gap equation
at $B=0$ for $\mu_5=0.15\,\mathrm{GeV}$, locating the temperature
peak of the scalar susceptibility derived from the same potential.
We vary $c_r$ until the ratio of this peak temperature to the one
at $\mu_5=0$ reproduces the adopted input,
$\mathcal R_\chi(0;c_r)=1.0057$.
The denominator is independent of $c_r$ because the added term
vanishes at $\mu_5=0$. D and MFDS coincide at $B=0$, so the
same input determines one coefficient for each vacuum regularization.
The additional lower and upper inputs define separate, nonstatistical
sensitivity scenarios. Their effects on the predicted model spread are
examined in Appendix~\ref{app:finite-source-input}. The adopted input yields
\begin{equation}
c_{\rm SC}=-0.07112,\qquad c_{\rm PV}=0.22479.
\label{eq:proof-match-coefficients}
\end{equation}
For each regularization, the reported coefficient is obtained with
Brent's method in the prescribed bracket $c_r\in[-0.2,0.5]$.

\section{Temperature response and model dependence in the original NJL model}
\label{sec:mu5B-phenomenology}
This section examines the original NJL treatments with $c_r=0$.
We first compare their finite-$\mu_5$ temperature responses and
identify the vacuum and thermal contributions that control them.
We then examine sensitivity to vacuum recalibration and use
magnetic inverse reconstruction as a separate diagnostic of model
dependence. The primary comparisons use the mass-inflection
criterion (1); a scalar-susceptibility peak check is identified
separately. Independent matching is considered in
Sec.~\ref{sec:phase-results}.

\subsection{Original temperature response in a magnetic field}
\label{sec:conditional-forward}
\label{sec:2x2-response}\label{sec:crossover-criterion-check}
We begin with the reference calibrations at $c_r=0$. Over the sampled
ranges $0\leq eB\leq0.4\,\mathrm{GeV}^2$ and
$0\leq\mu_5\leq0.30\,\mathrm{GeV}$, increasing $\mu_5$ lowers
the mass-inflection pseudocritical temperature in SC and raises it
in PV. This opposite $\mu_5$ dependence is already present at
$B=0$, where D and MFDS coincide within each vacuum regularization,
and persists at nonzero magnetic fields. Similar sensitivities of the
chiral-transition temperature to regularization and renormalization
prescriptions at finite $\mu_5$ have been discussed in
Refs.~\cite{Yu:2015hym,Farias:2016let,Ruggieri:2016cbq}. Within MFDS, SC-F and PV-F use the same ultraviolet-finite
magnetic contribution at a given $M,T,B,\mu_5$, although their
self-consistent equilibrium masses generally differ. The opposite
responses therefore cannot be attributed solely to the treatment
of the magnetic contribution.

\begin{figure*}[!htbp]
\centering
\includegraphics[width=.98\textwidth,height=.78\textheight,keepaspectratio]{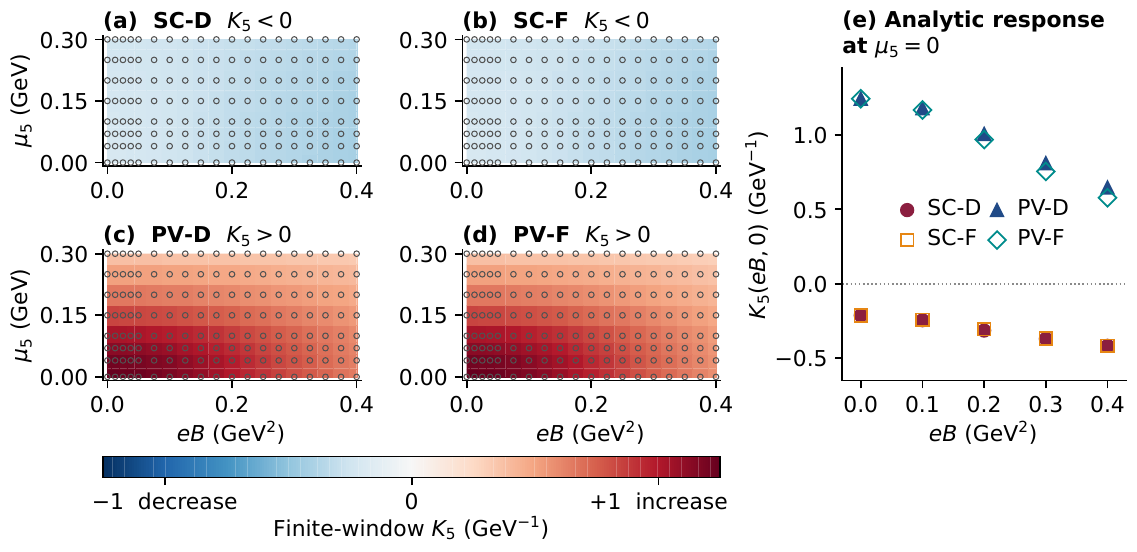}
\setlength{\abovecaptionskip}{2pt}\linespread{1}\selectfont
\caption{Original $c_r=0$ mass-inflection pseudocritical-temperature response at the reference vacuum calibrations. In (a)--(d), circles mark the $19\times8$ evaluated nodes per treatment. The magnetic-field spacing is $0.0125\,\mathrm{GeV}^2$ for $0\leq eB\leq0.05\,\mathrm{GeV}^2$ and $0.025\,\mathrm{GeV}^2$ for $0.05\leq eB\leq0.4\,\mathrm{GeV}^2$. The common color scale shows three-point finite-window slope estimates on the nonuniform $\mu_5^2$ grid. At each node, the slope is taken from a quadratic interpolant through three neighboring pseudocritical-temperature values on the $\mu_5^2$ grid, with one-sided stencils at the endpoints. These estimates are distinct from the analytic zero-source derivative $K_5(eB,0)$. Blue and red indicate negative and positive values. Panel (e) shows independent analytic derivatives at $\mu_5=0$ for the 20 reference-calibration anchors, covering $eB=0,0.1,0.2,0.3,0.4\,\mathrm{GeV}^2$ in each treatment. At zero magnetic field the D and F markers coincide within each baseline.}
\label{fig:response-matrix}
\end{figure*}

\begin{figure}[!htbp]
\centering
\includegraphics[width=.98\textwidth,keepaspectratio]{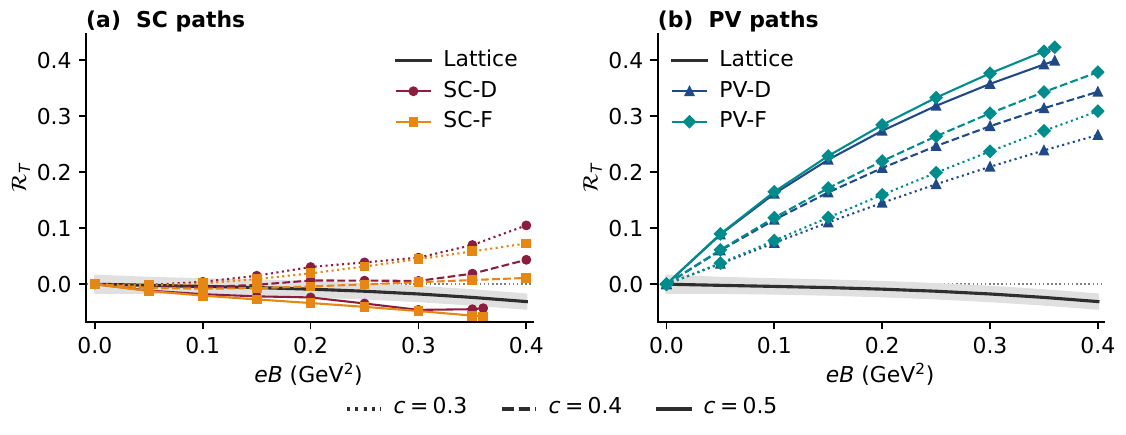}
\setlength{\abovecaptionskip}{2pt}\linespread{1}\selectfont
\caption{Normalized mass-inflection pseudocritical-temperature
trajectories for the original $c_r=0$ treatments along
$\mu_5=c\sqrt{eB}$, with $c=0.3,0.4,0.5$.
Panels (a) and (b) show the SC and PV treatments, respectively.
Each trajectory is normalized to its own $B=\mu_5=0$ reference
temperature. The black curve and gray band show the lattice-QCD
condensate-inflection result and its uncertainty.}
\label{fig:forward-paths}
\end{figure}

To illustrate this contrast, Fig.~\ref{fig:response-matrix} first
shows the finite-$\mu_5$ response at fixed magnetic fields, while
Fig.~\ref{fig:forward-paths} follows the corresponding
pseudocritical-temperature trajectories along the prescribed paths
$\mu_5(B)=c\sqrt{eB}$, with $c=0.3,0.4,0.5$, in the original
$c_r=0$ treatments. The dimensionless path parameter $c$ is distinct
from the additional coupling $c_r$. At fixed $eB$, increasing $c$
moves the SC predictions downward and the PV predictions upward.
Along each path both $eB$ and $\mu_5$ vary, so the trajectories
also include the magnetic-field dependence of each treatment's
$\mu_5=0$ reference background.

We normalize the mass-inflection pseudocritical-temperature change
for each treatment to its own reference value at $B=\mu_5=0$:
\begin{equation}
\mathcal R_T^{(s)}(B,\mu_5)=
\frac{T_{\rm pc}^{(1),s}(B,\mu_5)-T_{\rm pc}^{(1),s}(0,0)}{T_{\rm pc}^{(1),s}(0,0)}.
\label{eq:RT-model}
\end{equation}
Here $T_{\rm ref}^{(s)}=T_{\rm pc}^{(1),s}(0,0)$.
The sampled finite-window response and independent analytic
derivatives $K_5(eB,0)$ in Fig.~\ref{fig:response-matrix} are
distinct quantities, but their signs agree at the reference-calibration
anchors checked against independent finite-source trajectories. A representative calculation
with the scalar-susceptibility peak criterion also retains the
SC-negative/PV-positive contrast for these reference calibrations.
The dependence of the mass-inflection response on vacuum calibration
is examined in Sec.~\ref{sec:matching-robustness}.

\FloatBarrier

\subsection{Vacuum and thermal origins of the response}\label{sec:mechanism-results}
For the reference calculations with $c_r=0$, finite $\mu_5$ changes
the quark kernel at fixed constituent mass and, through the gap
equation, the equilibrium mass. Previous studies found that the
treatment of vacuum and thermal contributions can change the
pseudocritical-temperature trend \cite{Yu:2015hym,Farias:2016let}. Here the
complete physical thermal contribution is retained, allowing us to
examine the remaining sensitivity to the vacuum regularization.
At $x=\mu_5^2=0$, differentiation at fixed $M,T,eB$ gives
\begin{equation}
C_{\rm gap}=\left.\left(\frac{\partial R}{\partial x}\right)_{M,T,eB}
\right|_{x=0}=C_{\rm vac}+C_{\rm med},\qquad
\left.\left(\frac{\partial M_*}{\partial x}\right)_{T,eB}\right|_{x=0}
=-\frac{C_{\rm gap}}{R_M}.
\label{eq:Cgap-definition}
\end{equation}
Here $C_{\rm vac}$ and $C_{\rm med}$ are the respective vacuum and
thermal contributions to the derivative of the gap residual with
respect to $\mu_5^2$. They enter the equilibrium-mass response
through their sum, evaluated on the mass trajectory of each
treatment; $R_M>0$ on the selected stable branch.

\begin{figure*}[!htbp]
\centering
\includegraphics[width=.94\textwidth,height=.40\textheight,keepaspectratio]{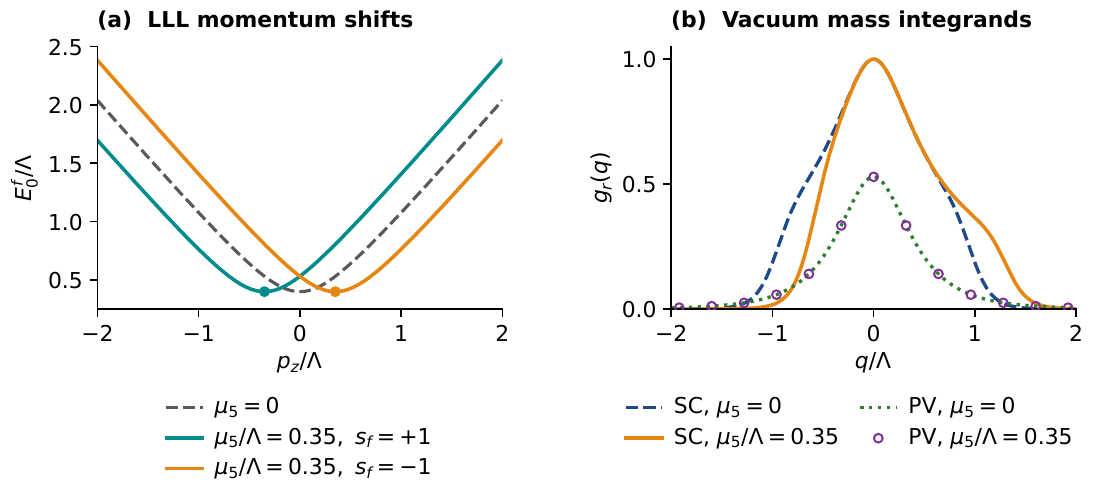}
\setlength{\abovecaptionskip}{2pt}\linespread{1}\selectfont
\caption{(a) Physical LLL dispersion relations for $s_f=\pm1$
at $\mu_5=0$ and $\mu_5/\Lambda=0.35$.
(b) SC and PV LLL vacuum mass integrands at fixed $M/\Lambda=0.4$,
plotted against the shifted momentum $q=p_z+\mu_5$
for $s_f=+1$:
$g_{\rm SC}(q;\mu_5)=w((q-\mu_5)^2/\Lambda^2)M/E_0(q)$ and
$g_{\rm PV}(q)=M\sum_iC_i/E_i(q)$, where
$w(u)=(1+u^5)^{-1}$ and $E_i(q)=\sqrt{M_i^2+q^2}$.
The SC integrand changes with $\mu_5$ through the soft-cutoff weight,
whereas the two PV curves coincide after the regulator-mass
combination and momentum shift.}
\label{fig:vacuum-dispersion-weights}
\end{figure*}

At $B=0$, the two positive-energy quark branches are
$E_\pm(k)=\sqrt{M^2+(k\pm\mu_5)^2}$, with $k=|\bm p|\geq0$.
For $\mu_5>0$, the lower dispersion branch is $E_-(k)$, whose
minimum is $M$ at $k=\mu_5$.
For Landau level $n$, this minimum is accessible when
$\mu_5^2\geq2n\mathcal B_f$, a condition automatically
satisfied for $n=0$.
To illustrate a longitudinal translation,
Fig.~\ref{fig:vacuum-dispersion-weights}(a) instead uses the LLL
energy of Eq.~\eqref{eq:LLL-energy}, with signed $p_z\in\mathbb R$.

For the even combination of opposite longitudinal shifts, set
$x=\mu_5^2$, $z=p^2$, and $F(z)=M/\sqrt{M^2+z}$. At fixed $M$,
the local source coefficient is
\begin{equation}
\left.\partial_x\frac{F((p+\mu_5)^2)+F((p-\mu_5)^2)}{2}\right|_{x=0}
=F_z+2zF_{zz}
=\frac{M(2z-M^2)}{2(M^2+z)^{5/2}}.
\label{eq:detail-vacuum-source-kernel}
\end{equation}
The coefficient changes sign at $|p|=M/\sqrt{2}$; its contribution
to $C_{\rm gap}$ also carries the minus sign from the gap residual.
For higher Landau levels the same branch-pair identity applies with
$p$ replaced by $S_{n,f}$.

Figure~\ref{fig:vacuum-dispersion-weights}(b) compares the complete
LLL vacuum mass integrands after setting $q=p_z+\mu_5$ for $s_f=+1$.
The SC weight is fixed at the unshifted momentum:
$w(p_z^2/\Lambda^2)=f_\Lambda(p_z^2)$. In the new variable it becomes
$w((q-\mu_5)^2/\Lambda^2)$, so the integral retains a source dependence
through the regulator-gradient term in Eq.~\eqref{eq:lll-smooth-window}.
This smooth weight imposes no hard integration boundary.
PV instead combines the mass kernels before integration.
The cancellations $\sum_iC_i=\sum_iC_iM_i^2=0$ leave an
$O(|q|^{-5})$ mass integrand, whose full-line integral is invariant
under the shift, as is the combined LLL vacuum potential in
Eq.~\eqref{eq:lll-pv-shift-invariance}.
This fixed-$M$, direct-LLL statement does not imply source independence
of the $B=0$ radial integral, higher Landau levels, the MFDS magnetic
difference, or the self-consistent response.

\begin{figure*}[!htbp]
\centering
\includegraphics[width=.96\textwidth,height=.65\textheight,keepaspectratio]{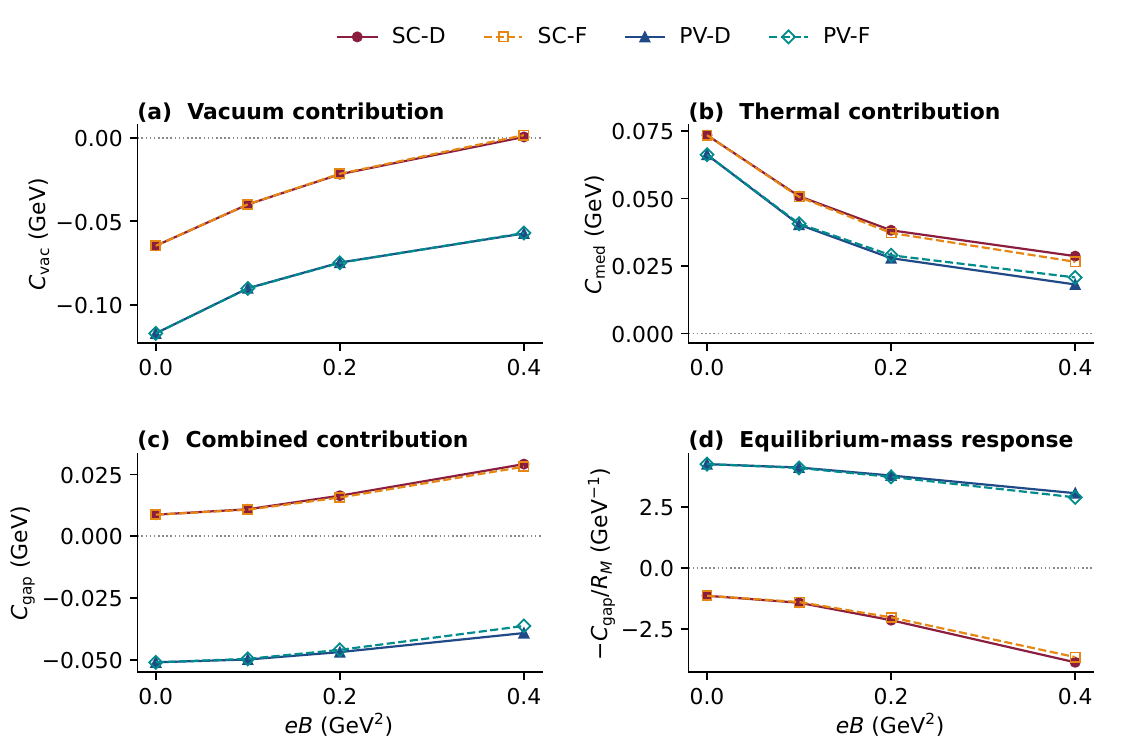}
\setlength{\abovecaptionskip}{2pt}\linespread{1}\selectfont
\caption{Vacuum (a), thermal (b), and combined (c) contributions to the $\mu_5^2$ derivative of the gap residual at the corresponding reference crossover temperatures, for $c_r=0$. The $\mu_5^2$ derivatives are evaluated by numerical finite differences at fixed $M,T$, whereas $R_M$ is obtained by analytic differentiation of the integrands with respect to $M$, followed by numerical integration. Panel (d) shows the resulting equilibrium-mass response $h=-C_{\rm gap}/R_M$. Each treatment is evaluated at its own equilibrium state; the temperature curvature of $h$ controls the local mass-inflection temperature shift.}
\label{fig:source-mechanism}
\end{figure*}

Both SC and PV use the same unregularized physical quark spectrum
in the thermal contribution. Its value nevertheless depends on the
equilibrium mass, which generally differs between the two vacuum
regularizations. At the reference crossover states shown in
Fig.~\ref{fig:source-mechanism}, the sampled thermal contributions
are positive. In PV, the negative vacuum term dominates, giving
$C_{\rm gap}<0$; in SC, their sum is positive. Thus the equilibrium-mass response
is positive in PV and negative in SC at these states. The local
pseudocritical-temperature response, however, depends on the
temperature curvature of this mass response, rather than its sign alone.

To relate this mass response to the local crossover-temperature
shift, one considers the selected smooth, decreasing mass trajectory.
At its nondegenerate temperature inflection,
$M_{TT}=0$ and $M_{TTT}>0$. Differentiating the inflection
condition gives an equivalent expression for $K_5$ as defined in
Eq.~\eqref{eq:K5-surface}:
\begin{equation}
K_5=-\left.
\frac{\bigl(\partial^3 M_*/\partial x\,\partial T^2\bigr)_{eB}}
     {\bigl(\partial^3 M_*/\partial T^3\bigr)_{x,eB}}
\right|_{T=T_{\rm pc}} .
\label{eq:K5-implicit-app}
\end{equation}
The implicit derivatives of the gap equation are given in
Appendix~\ref{app:source-matching-details}. At fixed $eB$, write
$h(T)=M_x(T,0)$. The same identity at $x=0$, with
Eq.~\eqref{eq:Cgap-definition} substituted, becomes
\begin{equation}
K_5(eB,0)=-\frac{h_{TT}}{M_{TTT}}
=\left.\frac1{M_{TTT}}\frac{d^2}{dT^2}
\left(\frac{C_{\rm vac}+C_{\rm med}}{R_M}\right)
\right|_{T=T_{\rm pc},\,x=0}.
\label{eq:detail-K5-mechanism}
\end{equation}
The temperature derivatives act on the full ratio along each
self-consistent mass trajectory. A temperature-independent correction
to the mass curve does not shift its temperature inflection at this
order; the local pseudocritical-temperature shift is instead
controlled by the temperature curvature $h_{TT}$. Appendix~\ref{app:detail-momentum-shifts} gives a complementary
LLL example of regulator-boundary effects.
Appendix Fig.~\ref{fig:polish-local-curvature} illustrates this
curvature dependence. At the sampled reference inflections,
$h_{TT}$ is positive for SC and negative for PV, while $M_{TTT}$
is positive in all four treatments. Consequently, $K_5$ is negative
for SC and positive for PV.

\subsection{Dependence on vacuum calibration}
\label{sec:matching-robustness}
We next ask whether the sign of the finite-$\mu_5$ temperature
response is fixed by the vacuum regularization alone. For this test, we construct vacuum parameter families fitted to
$(m_\pi,f_\pi,M_0)$ instead of
$(m_\pi,f_\pi,\langle\bar uu\rangle)$. Varying the vacuum
constituent mass $M_0$ determines corresponding sets of
$(m_0,G,\Lambda)$, with the condensate treated as an output.
The families retain $c_r=0$ and are distinct from the independent
finite-$\mu_5$ matching of Sec.~\ref{sec:source-input}.

\begin{figure*}[!htbp]
\centering
\includegraphics[width=.98\textwidth,height=.78\textheight,keepaspectratio]{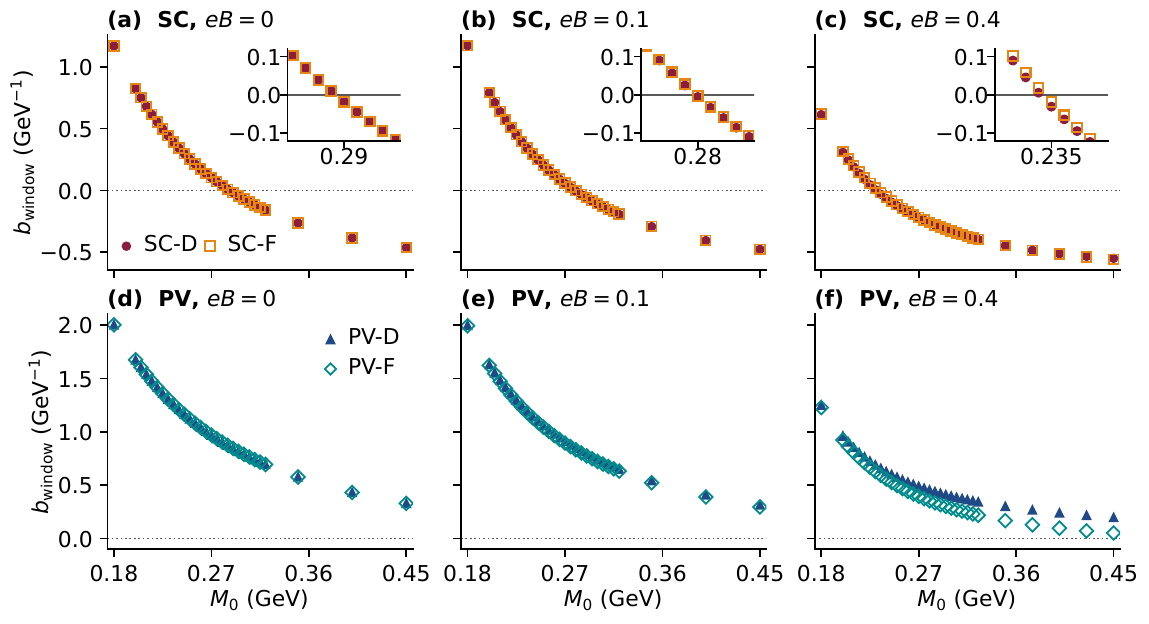}
\setlength{\abovecaptionskip}{2pt}\linespread{1}\selectfont
\caption{Dependence on vacuum recalibration through $M_0$ at $eB=0,0.1,0.4\,\mathrm{GeV}^2$, with $c_r=0$. Points give $b_{\rm window}$ from the equal-weight fit $T_{\rm pc}^{(1)}=a+b_{\rm window}\mu_5^2+d\mu_5^4$ at $\mu_5=0,0.04,0.07,0.10\,\mathrm{GeV}$; this finite-window coefficient is not assumed equal to the local derivative $K_5(eB,0)$. Insets resolve sampled SC sign changes near zero, while the examined PV samples remain positive. D and F coincide at zero magnetic field. The displayed range ends at $M_0=0.45\,\mathrm{GeV}$; the full sampled family extends to $0.54\,\mathrm{GeV}$. }
\label{fig:matching-family}
\end{figure*}

Figure~\ref{fig:matching-family} shows the finite-window coefficient
$b_{\rm window}$ obtained by fitting
$T_{\rm pc}^{(1)}=a+b_{\rm window}\mu_5^2+d\mu_5^4$ in the
specified small-$\mu_5$ interval. The examined SC families contain
both positive and negative responses, while the sampled PV
calibrations retain positive responses. For 31 near-zero responses,
independent local derivatives agree with the finite-window signs.
Narrowing the source windows produces no sign reversal under the
tested refinements.

The reference and recalibrated families show that conventional
vacuum inputs do not determine a unique finite-$\mu_5$
temperature response. The positive response of the examined PV
families is established only over their sampled parameter range.

\subsection{Magnetic inverse reconstruction as a diagnostic}
\label{sec:conditional-inverse}\label{sec:conditional-family}
We use the normalized mass-inflection temperature change
$\mathcal R_T^{(s)}$ defined in Eq.~\eqref{eq:RT-model}.
When the normalized lattice-QCD trend is used below as an inverse
diagnostic, its normalization is independently defined at $B=0$.
We therefore compare relative changes, rather than fit the model's
absolute pseudocritical temperature to the lattice value.

The normalized lattice-QCD crossover-temperature trend also allows
an inverse test of the original responses. For each treatment with
$c_r=0$, we ask whether a magnetic-field-dependent effective chiral
chemical potential reaches the central target within the explored
interval:
\begin{equation}
\mathcal R_T^{(s)}(B,\mu_{5,\mathrm{eff}}^{(s)}(B))
=\mathcal R_T^{\rm lat}(B),\qquad
0\leq\mu_{5,\mathrm{eff}}^{(s)}\leq0.30\,\mathrm{GeV}.
\label{eq:mu5eff-inverse}
\end{equation}
This condition is a diagnostic, not an additional matching
prescription. It combines the $\mu_5=0$ magnetic background with
the finite-$\mu_5$ response, and a root need not exist over the
checked range.

The magnetic target for the mass-inflection criterion (1) is taken
from the continuum-limit condensate-inflection curve and its band in
the upper-left panel of Fig.~9 of Ref.~\cite{Bali:2011qj}.
The target and band are normalized to the central curve at zero
magnetic field ($B=0$). The digitized band is used only for sampled
overlap checks, without assuming covariance information. Roots of
Eq.~\eqref{eq:mu5eff-inverse} are checked by direct substitution;
band overlap is distinguished from a central-target root.

\begin{figure}[!htbp]
\centering
\includegraphics[width=.98\textwidth,keepaspectratio]{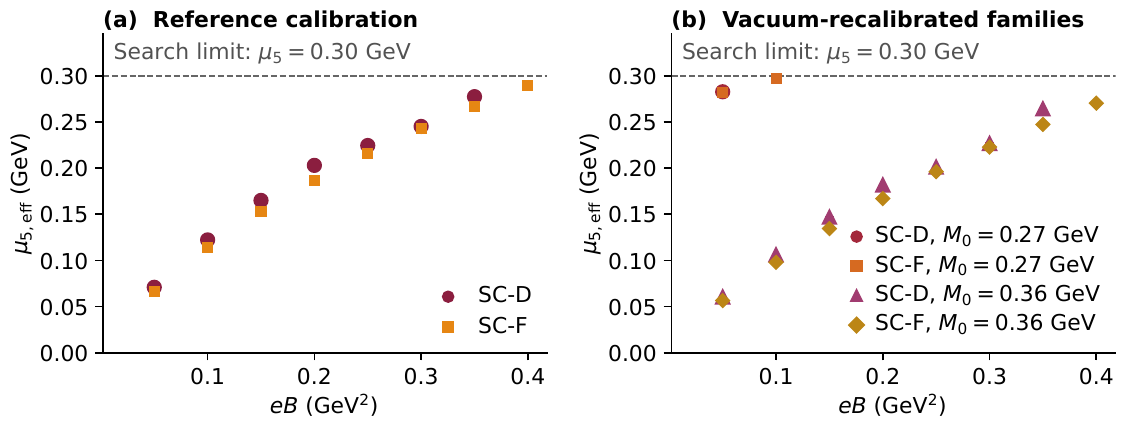}
\setlength{\abovecaptionskip}{2pt}\linespread{1}\selectfont
\caption{Central-target inverse roots $\mu_{5,\mathrm{eff}}$ in the
original $c_r=0$ treatments, using the mass-inflection criterion (1).
Panel (a) uses the reference vacuum parameters; panel (b) compares
SC-D and SC-F after vacuum recalibration with
$M_0=0.27$ and $0.36\,\mathrm{GeV}$. No central-target roots were found for the examined PV
families or the $M_0=0.18\,\mathrm{GeV}$ family within the checked
interval. The dashed line marks the computational search limit
$\mu_5=0.30\,\mathrm{GeV}$, not a physical critical boundary.}
\label{fig:inverse-roots}
\end{figure}

Figure~\ref{fig:inverse-roots} compares the resolved central-target
roots for the reference calibrations and selected vacuum-recalibrated
families. At the reference vacuum calibrations, several sampled SC
cases admit resolved central-target roots, whereas the PV treatments
do not reach the target at $eB=0.4\,\mathrm{GeV}^2$.
At that field, SC-F has
$\mu_{5,\mathrm{eff}}\simeq0.29\,\mathrm{GeV}$, while SC-D has no
central-target root in the searched interval. The number and
positions of the roots also depend on the vacuum recalibration
considered above.

Together, the reference calculations, vacuum-recalibration families,
and inverse diagnostic show that conventional vacuum inputs alone do
not select a common mass-inflection temperature response across SC and
PV at finite $\mu_5$. Although the examined SC families contain both
response signs, these variations do not by themselves establish a
common prediction with PV. This motivates fixing the additional
coefficient $c_r$ from an independent finite-$\mu_5$ observable and
testing whether the spread among treatments is reduced.

\FloatBarrier
\section{Effects of independent finite-\texorpdfstring{$\mu_5$}{mu5} matching}
\label{sec:phase-results}
We now use the independently matched coefficients
$c_{\rm SC}=-0.07112$ and $c_{\rm PV}=0.22479$ determined in
Sec.~\ref{sec:source-input} from the common $B=0$ scalar
peak-temperature ratio $y_0=1.0057$. Each coefficient is shared
by D and MFDS within its vacuum regularization and held fixed
at finite magnetic fields. We first compare the induced
mass-inflection temperature changes and then the scalar
peak-temperature ratios. Neither comparison involves additional
fitting at nonzero magnetic field.

\subsection{Response changes after independent matching}
\label{sec:matched-response}
To separate the response to finite $\mu_5$ from the magnetic
background of each treatment, we define, for each treatment at
a fixed $B$ and its prescribed $c_r$,
\begin{equation}
\Delta\mathcal R_T^{(s)}(\mu_5)
=\mathcal R_T^{(s)}(B,\mu_5;c_r)-\mathcal R_T^{(s)}(B,0).
\label{eq:matched-temperature-response}
\end{equation}
The arguments $B$ and $c_r$ are suppressed on the left-hand side.
Here $\mathcal R_T^{(s)}$ uses the mass-inflection criterion (1)
and the normalization $T_{\rm ref}^{(s)}=T_{\rm pc}^{(1),s}(0,0)$
of Eq.~\eqref{eq:RT-model}. The added interaction vanishes at
$\mu_5=0$, so the subtracted magnetic background is unchanged
by matching.

A local analytic relation clarifies how the additional
interaction modifies the temperature response.
If $R_0(M,T,B,x)$ denotes the original gap residual, the matched
residual is
$R_c=R_0+2c_rx(M-m_0)$.
At $x=0$, the two residuals define the same equilibrium mass
$M_*(T,0)$ and the same $R_{0,M}$ along that trajectory. Implicit
differentiation gives
\begin{align}
h_c(T)&\equiv
\left.\frac{\partial M_*^{(c)}(T,x)}{\partial x}\right|_{x=0}
=-\frac{R_{0,x}+2c_r\bigl(M_*(T,0)-m_0\bigr)}{R_{0,M}},\notag\\
h_c(T)-h_0(T)&=-\frac{2c_r\bigl(M_*(T,0)-m_0\bigr)}{R_{0,M}}.
\label{eq:matched-mass-response}
\end{align}
All derivatives on the right are evaluated along the common
$\mu_5=0$ equilibrium trajectory. On the selected stable branch,
$R_{0,M}>0$ and $M_*>m_0$, so $c_{\rm SC}<0$ increases $h$
relative to its original value, whereas $c_{\rm PV}>0$ decreases it.
This can reduce the SC--PV contrast in the mass response. The
corresponding local mass-inflection temperature response depends
on the temperature curvature of $h$, as derived in
Appendix~\ref{app:detail-local-derivatives}
[Eq.~\eqref{eq:matched-local-K5-shift}]. The full finite-$\mu_5$
responses presented below are obtained from the extended gap equation.

\begin{figure*}[!htbp]
\centering
\includegraphics[width=.98\textwidth,height=.78\textheight,keepaspectratio]{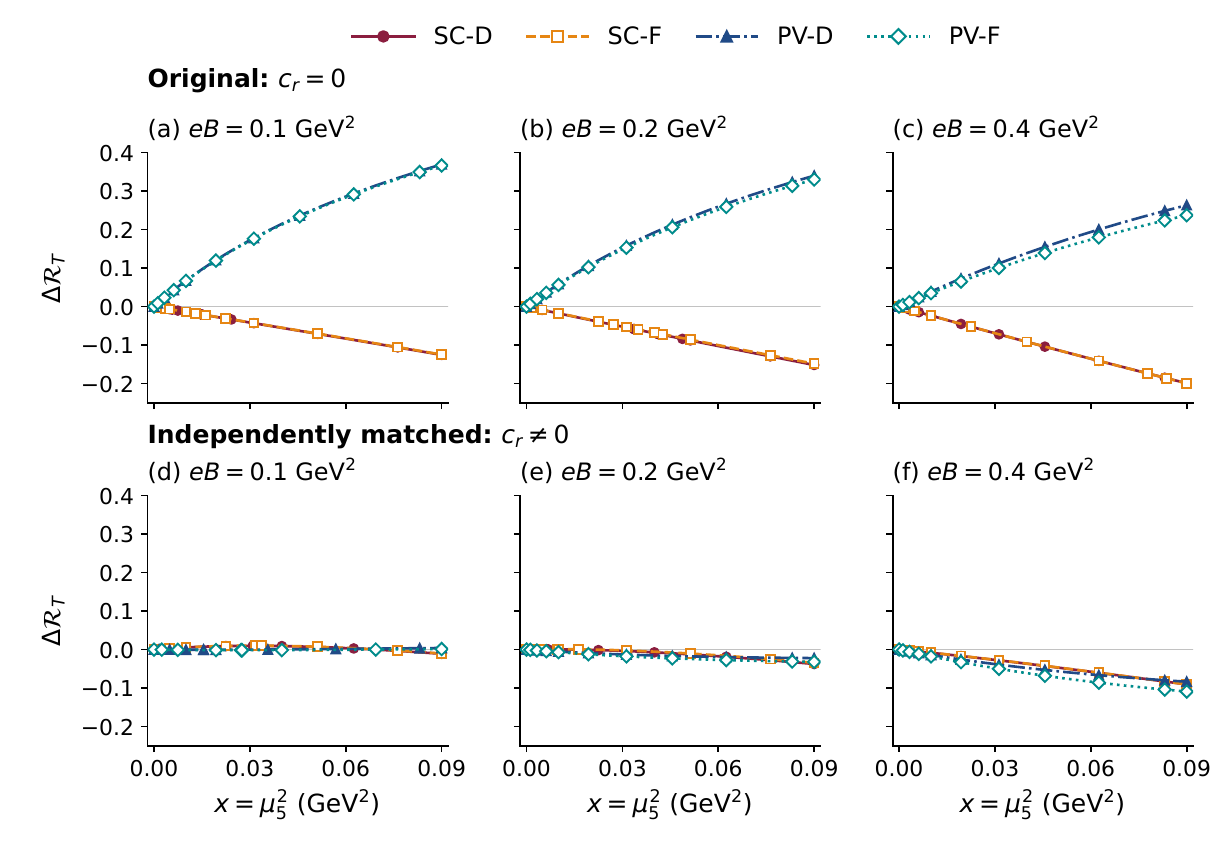}
\setlength{\abovecaptionskip}{2pt}\linespread{1}\selectfont
\caption{Comparison of finite-$\mu_5$-induced changes in the normalized
mass-inflection pseudocritical temperature among SC-D, SC-F, PV-D,
and PV-F. The upper panels use $c_r=0$; the lower panels use
independently matched $c_{\rm SC}$ and $c_{\rm PV}$ from the
common $B=0$ input $y_0=1.0057$, held fixed at nonzero magnetic fields.
Columns correspond to $eB=0.1,0.2,0.4\,\mathrm{GeV}^2$.
For each treatment, the value at $\mu_5=0$ is subtracted;
normalization remains $T_{\rm pc}^{(1),s}(0,0)$.
All panels use identical scales.}
\label{fig:constrained-response}
\end{figure*}

As shown in Fig.~\ref{fig:constrained-response}, the original
reference SC curves decrease with increasing $\mu_5$, whereas the
PV curves increase. After matching, the
four induced temperature changes are substantially closer over
the sampled range. Their shapes are not identical: at
$eB=0.1\,\mathrm{GeV}^2$, the matched SC curves first rise and
then fall, while the PV curves first fall and then rise. At
$eB=0.2\,\mathrm{GeV}^2$, SC retains a small initial rise,
whereas PV decreases. At $eB=0.4\,\mathrm{GeV}^2$, all four matched curves show a
decreasing trend with increasing $\mu_5$.

For a pointwise comparison, define
$S_{\Delta T}=\max_s\Delta\mathcal R_T^{(s)}(\mu_5)
-\min_s\Delta\mathcal R_T^{(s)}(\mu_5)$ at fixed $B$ and $\mu_5$.
Table~\ref{tab:matched-response-spreads} reports this spread at
$\mu_5=0.15$ and $0.30\,\mathrm{GeV}$. At the three finite magnetic fields, the reductions are by factors
of approximately $7.5$--$15.3$ at $\mu_5=0.15\,\mathrm{GeV}$ and
$18.1$--$32.6$ at $\mu_5=0.30\,\mathrm{GeV}$. The spread is also smaller at every other common nonzero sampled
$\mu_5$ value. The $\mu_5=0.30\,\mathrm{GeV}$ results provide
an additional finite-source prediction beyond the matching point
at $\mu_5=0.15\,\mathrm{GeV}$.

\begin{table}[!htbp]
\caption{Spread $S_{\Delta T}$ of the normalized finite-$\mu_5$
mass-inflection temperature response shown in
Fig.~\ref{fig:constrained-response}. ``Before'' denotes $c_r=0$;
``After'' denotes the coefficients matched at $B=0$.
Contraction factors use unrounded values.}
\label{tab:matched-response-spreads}
\begin{ruledtabular}
\begin{tabular}{ccrrr}
$\mu_5$ (GeV) & $eB$ (GeV$^2$)
& Before & After & Before/after \\
\hline
0.15 & 0.1 & 0.16779 & 0.01094 & 15.34 \\
     & 0.2 & 0.15979 & 0.01354 & 11.81 \\
     & 0.4 & 0.13563 & 0.01819 &  7.46 \\
\hline
0.30 & 0.1 & 0.49476 & 0.01520 & 32.55 \\
     & 0.2 & 0.49162 & 0.01534 & 32.06 \\
     & 0.4 & 0.46272 & 0.02557 & 18.10
\end{tabular}
\end{ruledtabular}
\end{table}

For a smooth equilibrium branch at fixed $B$ and $c_r$, the full
finite-$\mu_5$ response satisfies
\begin{equation}
\Delta\mathcal R_T^{(s)}(\mu_5)
={}\frac{1}{T_{\rm ref}^{(s)}}
\int_0^{\mu_5^2}dx\,
K_5^{(s)}(eB,\sqrt{x};c_r).
\label{eq:constrained-finite-shift}
\end{equation}
The subtracted $\mu_5=0$ magnetic background is unchanged by
matching, while the integral gives the accumulated finite-$\mu_5$
response shown in
Fig.~\ref{fig:constrained-response}. The local derivative
$K_5(eB,0)$ alone does not determine this finite-interval change.

 \subsection{Reduction of scalar peak-temperature-ratio spread}
\label{sec:proof-matching}
\begin{figure*}[!htbp]
\centering
\includegraphics[width=.98\textwidth,height=.78\textheight,keepaspectratio]{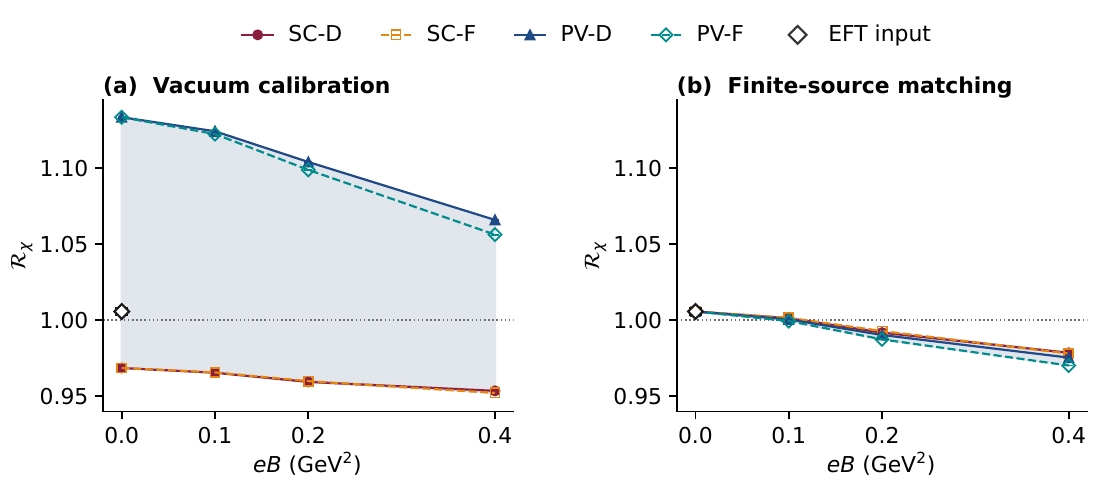}
\setlength{\abovecaptionskip}{2pt}\linespread{1}\selectfont
\caption{Scalar peak-temperature ratio at $\mu_5=0.15\,\mathrm{GeV}$, normalized to the peak at $\mu_5=0$ at the same magnetic field, for the original $c_r=0$ (a) and matched-$c_r$ (b) treatments, on the same scale. The diamond and its bar at zero magnetic field give the effective-theory input $y_0=1.0057$ and its nonstatistical interval~\cite{GomezNicola:2023ghi}. Shading spans the four central values. Table~\ref{tab:matching-spreads} gives the absolute spreads and their ratios at finite magnetic fields. Table~\ref{tab:polish-spreads} in Appendix~\ref{app:finite-source-input} gives the discrete common-input sensitivity.}
\label{fig:proof-matching}
\end{figure*}

We next test the scalar peak-temperature ratio $\mathcal R_\chi^{(s)}$
of Eq.~\eqref{eq:Rchi-proof} at the finite magnetic fields
$eB=0.1,0.2,0.4\,\mathrm{GeV}^2$ after matching at $B=0$
(Fig.~\ref{fig:proof-matching}). The four-treatment spread
$S=\max_s\mathcal R_\chi^{(s)}-\min_s\mathcal R_\chi^{(s)}$
decreases by factors of approximately $74$, $27$, and $14$ at these
magnetic fields, respectively. Table~\ref{tab:matching-spreads} reports
the corresponding absolute spreads and contraction factors calculated
from unrounded values. The common scalar peak-temperature ratio
$y_0=1.0057$ is imposed only at $B=0$, with $c_r$ held fixed at nonzero
magnetic fields. The observed reduction in spread therefore provides
a prediction beyond the matching condition, rather than a direct
consequence of it. The sensitivity to the common matching input is
examined in Appendix~\ref{app:finite-source-input} and
Table~\ref{tab:polish-spreads}.

\begin{table}[!htbp]
\caption{Model spread $S$ of the scalar peak ratio at
$\mu_5=0.15\,\mathrm{GeV}$ before and after matching $c_r$ with
$y_0=1.0057$, at finite magnetic fields. The contraction
factors use the unrounded spreads.}
\label{tab:matching-spreads}
\begin{ruledtabular}
\begin{tabular}{crrr}
$eB$ (GeV$^2$) & Before & After & Before/after\\
\hline
0.1 & 0.15872 & 0.00214 & 74.24\\
0.2 & 0.14457 & 0.00544 & 26.58\\
0.4 & 0.11373 & 0.00835 & 13.62
\end{tabular}
\end{ruledtabular}
\end{table}

The additional interaction vanishes at $\mu_5=0$, leaving the
corresponding equilibrium mass trajectories and reference peak
temperatures unchanged. At finite $\mu_5$, it modifies the gap
equation and hence the temperature dependence of the scalar
susceptibility along the new equilibrium trajectory. The reported
peak ratios are obtained from these full finite-$\mu_5$ equilibrium
calculations.

The spread reductions in Tables~\ref{tab:matched-response-spreads}
and~\ref{tab:matching-spreads} concern different observables:
the mass-inflection criterion (1) and the scalar peak criterion (2),
respectively. Their factors cannot be interchanged, and neither
implies equal absolute crossover temperatures, statistical
uncertainties, or agreement with QCD.

\section{Discussion and conclusions}\label{sec:discussion-conclusions}
\begingroup\widowpenalty=10000
We have investigated whether independent information at finite chiral chemical
potential $\mu_5$ can constrain the magnetic-field response of the chiral
crossover within the Nambu--Jona-Lasinio model. Conventional vacuum
calibration leaves the coefficient $c_r$ of the additional interaction
$\delta\Omega_r=c_r\mu_5^2(M-m_0)^2$ undetermined. Using the same $B=0$
scalar-susceptibility peak-temperature ratio predicted by unitarized chiral
perturbation theory, we obtain $c_{\rm SC}=-0.07112$ and $c_{\rm PV}=0.22479$.
Each coefficient is shared by both magnetic-field treatments within its vacuum
regularization and held fixed at nonzero magnetic fields.
\par\endgroup

In the original $c_r=0$ model, increasing $\mu_5$ lowers the mass-inflection
pseudocritical temperature in the reference SC treatments but raises it in PV
over the sampled domain. The examined SC vacuum-recalibration families contain
both response signs, whereas the sampled PV families retain a positive
response. The vacuum and thermal contributions enter the equilibrium-mass
response through the gap equation, while the temperature curvature of that
response controls the local mass-inflection temperature shift. Across the
examined reference and recalibrated cases, vacuum calibration alone does
not select a common SC/PV finite-$\mu_5$ response. This motivates the
additional finite-source constraint implemented through the matched $c_r$.

With the independently matched coefficients, the spread among the four
treatments in the normalized finite-$\mu_5$-induced mass-inflection
temperature change decreases by factors of approximately $7.5$--$15.3$ at
$\mu_5=0.15\,\mathrm{GeV}$ and $18.1$--$32.6$ at $\mu_5=0.30\,\mathrm{GeV}$,
across the finite magnetic fields $eB=0.1,\,0.2,\,0.4\,\mathrm{GeV}^2$. As
matching uses the scalar-susceptibility peak-temperature ratio, this reduction
tests a different crossover criterion. The response curves nevertheless retain
distinguishable $\mu_5$ dependences, so the reduced spread does not imply
identical response functions.

For the distinct scalar-susceptibility peak-temperature ratio at
$\mu_5=0.15\,\mathrm{GeV}$, the four-treatment spread decreases by factors of
approximately $74$, $27$, and $14$ at $eB=0.1,\,0.2,\,0.4\,\mathrm{GeV}^2$,
respectively. The original SC--PV difference in the finite-$\mu_5$
response is already present at $eB=0$, where D and MFDS coincide
within each vacuum regularization. This difference therefore cannot
be attributed to the magnetic field alone. By determining $c_r$
separately for SC and PV from the same $eB=0$ peak-temperature ratio,
the predicted finite-$\mu_5$ responses become substantially closer,
and this reduction in treatment dependence persists at the three
nonzero magnetic fields without further matching. Because the additional
interaction vanishes at $\mu_5=0$, this improvement concerns the
finite-$\mu_5$-induced temperature response, while each treatment
retains its original magnetic-field dependence of the absolute
pseudocritical temperature.

Further independent determinations of finite-$\mu_5$ response observables,
particularly at nonzero magnetic fields, would test the predictive scope of
the matching condition. Extensions beyond the single-operator interaction~\cite{Espriu:2020dge},
constrained by additional information~\cite{Chao:2022bbv,GomezNicola:2023ghi}, would help assess the robustness of the
reduced model spread. Mesonic correlations~\cite{Mao:2016fha,Chao:2026fdq} and axial-vector interactions~\cite{Zhang:2012rv} offer
complementary directions, but require separate consistent calculations. The
generation and relaxation of axial imbalance also lie outside the present
quasistatic treatment and call for a dynamical framework~\cite{Ruggieri:2016asg,Lin:2018nxj}.

\section*{Data availability}
Supporting data are retained by the authors but are not currently
deposited in a public repository. Inquiries may be directed to the authors.

\section*{Acknowledgments}
This work is supported by the National Natural Science
Foundation of China (NSFC) under Grant No. 12365020.

\appendix
\section{Landau-level construction of the propagator and static trace}
\label{app:propagator}
We construct the propagator in a basis that resolves its Dirac
structure and transverse orbital states. This representation fixes
the lowest-Landau-level (LLL) projection, orbital degeneracy, and
scalar weight entering the gap equation consistently. The transverse
spectral construction follows the method of Appendix A of
Ref.~\cite{Gorbar:2008hu}, extended here to the longitudinal
momentum and axial source of the present $3+1$-dimensional model.
The mass $M$ and chiral chemical potential $\mu_5$ are spatially
homogeneous. Ultraviolet regularization is introduced only after
identifying the scalar vacuum contribution.

\subsection{Orbital wave functions and spin-resolved Landau levels}
\label{appA:orbital-states}
We use $\hbar=c=1$, $g^{\mu\nu}=\operatorname{diag}(1,-1,-1,-1)$,
and $\gamma^5=i\gamma^0\gamma^1\gamma^2\gamma^3$.
For a flavor of charge $q_f$, choose the Landau gauge
$\bm A=(0,Bx,0)$ and the momentum convention of the main text,
\begin{align}
\pi_{f,x}&=-i\partial_x,\qquad
\pi_{f,y}=-i\partial_y+q_fBx,\qquad
\pi_{f,z}=p_z,\notag\\
[\pi_{f,x},\pi_{f,y}]&=-iq_fB,\qquad
\mathcal B_f=|q_fB|,\qquad
s_f=\operatorname{sgn}(q_fB).
\label{eq:appA-momentum-convention}
\end{align}
The formulas below refer to $\mathcal B_f>0$.  After Fourier
transformation in time and $z$, the Dirac operator is
\begin{equation}
\mathcal D_f={}\omega\gamma^0-p_z\gamma^3
-\gamma^1\pi_{f,x}-\gamma^2\pi_{f,y}
+\mu_5\gamma^0\gamma^5-M.
\label{eq:appA-dirac-operator}
\end{equation}
We define the Feynman propagator by $S_f=i\mathcal D_f^{-1}$.
To distinguish spacetime and transverse coordinates, write
$X=(t,x,y,z)$ and $X'=(t',x',y',z')$, with
$\bm\rho=(x,y)$ and $\bm\rho'=(x',y')$. The mixed representation is
\begin{equation}
S_f(X,X')=\int\frac{d\omega\,dp_z}{(2\pi)^2}
e^{-i\omega(t-t')+ip_z(z-z')}
S_f(\omega,p_z;\bm\rho,\bm\rho'),
\label{eq:appA-mixed-representation}
\end{equation}
The pole assignment will be specified after the physical denominators
have been separated.

For a fixed momentum $k$ conjugate to $y$, introduce
\begin{equation}
\ell_f=\mathcal B_f^{-1/2},\qquad
\zeta_{f,k}=\frac{x+k/(q_fB)}{\ell_f}.
\label{eq:appA-orbital-coordinate}
\end{equation}
Then $\pi_{f,y}=s_f\sqrt{\mathcal B_f}\,\zeta_{f,k}$ and
$\pi_{f,x}=-i\sqrt{\mathcal B_f}\,\partial_{\zeta_{f,k}}$.
The transverse kinetic operator is a harmonic oscillator:
\begin{equation}
\pi_{f,\perp}^2
=\mathcal B_f(-\partial_\zeta^2+\zeta^2)
=\mathcal B_f(2a^\dagger a+1),\qquad
a=\frac{\zeta+\partial_\zeta}{\sqrt2}.
\label{eq:appA-oscillator}
\end{equation}
Its normalized orbital modes are
\begin{equation}
\psi_{mk}^f(\bm\rho)
=\frac{e^{iky}}{\sqrt{2\pi\ell_f}}
\frac{e^{-\zeta_{f,k}^2/2}H_m(\zeta_{f,k})}
{\sqrt{2^m m!\sqrt\pi}},\qquad m=0,1,\ldots,
\label{eq:appA-orbital-modes}
\end{equation}
with eigenvalues $(2m+1)\mathcal B_f$.
The orbital center is $x_c=-k/(q_fB)$, so $k$ labels the continuous
guiding-center degeneracy, not an additional transverse excitation.  Our normalization gives
\begin{align}
\int d^2\rho\,
\psi_{mk}^{f*}(\bm\rho)\psi_{m'k'}^f(\bm\rho)
&=\delta_{mm'}\delta(k-k'),\notag\\
\sum_{m=0}^{\infty}\int_{-\infty}^{\infty}dk\,
\psi_{mk}^f(\bm\rho)\psi_{mk}^{f*}(\bm\rho')
&=\delta^{(2)}(\bm\rho-\bm\rho').
\label{eq:appA-orbital-completeness}
\end{align}
The factor $1/\sqrt{2\pi}$ in Eq.~\eqref{eq:appA-orbital-modes}
is already included; the guiding-center measure in
Eq.~\eqref{eq:appA-orbital-completeness} is therefore $dk$.

The orbital index $m$ differs from the physical Landau-level index.
With $\Sigma^j=\gamma^5\gamma^0\gamma^j$, define
\begin{equation}
h_f=\bm\Sigma\cdot\bm\pi_f,\qquad
\Sigma^3=i\gamma^1\gamma^2,\qquad
P_\pm^f=\frac{1\pm s_f\Sigma^3}{2}.
\label{eq:appA-spin-projectors}
\end{equation}
The momentum commutator in
Eq.~\eqref{eq:appA-momentum-convention} fixes the spin term:
\begin{equation}
h_f^2=p_z^2+\pi_{f,\perp}^2+q_fB\Sigma^3.
\label{eq:appA-h-square}
\end{equation}
For a spin eigenvector with $\Sigma^3\chi_\sigma=\sigma\chi_\sigma$,
the transverse eigenvalue is
\begin{equation}
\mathcal B_f(2m+1+s_f\sigma)=2n\mathcal B_f,
\qquad n=m+\frac{1+s_f\sigma}{2}.
\label{eq:appA-level-reindexing}
\end{equation}
Consequently, the physical level $n$ combines an orbital of index
$m=n$ in $P_-^f$ with one of index $m=n-1$ in $P_+^f$.
The second orbital exists only for $n\geq1$.
At $n=0$ the unavailable $m=-1$ state is absent, leaving only
$\psi_{0k}^fP_-^f$.  The LLL spin projection therefore follows from the oscillator
spectrum and the nonnegative orbital index; it is not an independent
degeneracy rule imposed on the propagator.
We set $\psi_{-1,k}^f=0$ below to express both cases uniformly.

\subsection{Guiding-center sum and the Schwinger phase}
\label{appA:orbital-kernel}
The coordinate kernel of a fixed orbital follows directly from
Eq.~\eqref{eq:appA-orbital-modes}.  Let
$\bm r=\bm\rho-\bm\rho'$ and
$\xi_f=\mathcal B_f\bm r^2/2$.  The guiding-center integral is
\begin{align}
\mathcal I_m^f(\bm\rho,\bm\rho')
&\equiv\int_{-\infty}^{\infty}dk\,
\psi_{mk}^f(\bm\rho)\psi_{mk}^{f*}(\bm\rho')=\frac{\mathcal B_f}{2\pi}
e^{i\Phi_f(\bm\rho,\bm\rho')}
e^{-\xi_f/2}L_m(\xi_f),
\label{eq:appA-guiding-center-kernel}\\
\Phi_f(\bm\rho,\bm\rho')
&=-\frac{q_fB}{2}(x+x')(y-y')
=-q_f\int_{\bm\rho'}^{\bm\rho}\bm A\cdot d\bm l.
\label{eq:appA-schwinger-phase}
\end{align}
The path in the last expression is the straight segment. The
integral over the guiding center is evaluated by completing the
Gaussian square in $k$ and applying the Hermite identity
\begin{equation}
\int_{-\infty}^{\infty}du\,e^{-u^2}H_m(u+a)H_m(u+b)
=2^m m!\sqrt\pi\,L_m(-2ab).
\label{eq:appA-hermite-identity}
\end{equation}
The resulting Laguerre polynomial encodes the overlap of the two
transverse oscillator wave functions.

The projector kernel for a physical Landau level is therefore
\begin{align}
\Pi_n^f(\bm\rho,\bm\rho')
&=\mathcal I_n^f(\bm\rho,\bm\rho')P_-^f
+\mathcal I_{n-1}^f(\bm\rho,\bm\rho')P_+^f\notag\\
&=\frac{\mathcal B_f}{2\pi}e^{i\Phi_f}e^{-\xi_f/2}
\big[L_n(\xi_f)P_-^f+L_{n-1}(\xi_f)P_+^f\big],
\label{eq:appA-landau-projector}
\end{align}
where $\mathcal I_{-1}^f=L_{-1}=0$.
Equation~\eqref{eq:appA-orbital-completeness} implies
$\Pi_n^f\Pi_{n'}^f=\delta_{nn'}\Pi_n^f$ and
$\sum_n\Pi_n^f=1$, with operator products understood to include
the intermediate transverse coordinate integral.
The $L_n$ and $L_{n-1}$ terms in
Eq.~\eqref{eq:appA-landau-projector} retain the spin--orbital pairing
of Eq.~\eqref{eq:appA-level-reindexing}.

All dependence on the mean transverse position is carried by
$\Phi_f$.  The remaining kernel depends only on $\bm r$.
This is the Schwinger-phase factorization in a uniform magnetic
field.  The phase retains the sign of $q_fB$; replacing it by
$\mathcal B_f$ would be incorrect for the oppositely charged flavor.  At coincident points the phase
is unity and $L_m(0)=1$, so
\begin{equation}
\Pi_n^f(\bm\rho,\bm\rho)
=\frac{\mathcal B_f}{2\pi}
\big[P_-^f+\theta(n-1)P_+^f\big].
\label{eq:appA-coincident-projector}
\end{equation}
Here $\theta(n-1)$ is one for $n\geq1$ and zero for $n=0$.
Thus the orbital density $\mathcal B_f/(2\pi)$ and the missing
LLL spin sector are fixed before the Dirac-matrix inversion.

\subsection{Ordered inversion and the momentum-space propagator}
\label{appA:ordered-inverse}
To obtain an explicit transverse-momentum representation, it is useful
to rationalize the matrix inverse before evaluating the orbital kernel.
For this calculation write
\begin{equation}
A=\omega\gamma^0-p_z\gamma^3,\qquad
C=\mu_5\gamma^0\gamma^5-M,\qquad
P=\pi_{f,x}\gamma^1+\pi_{f,y}\gamma^2,
\label{eq:appA-ACP}
\end{equation}
so that $\mathcal D_f=A+C-P$.
Right multiplication preserves the operator ordering needed for the
subsequent coordinate-space evaluation.
The identities $\{A,P\}=0$ and $[C,P]=0$, together with
\begin{equation}
P^2=-\pi_{f,\perp}^2-iq_fB\gamma^1\gamma^2,
\label{eq:appA-transverse-square}
\end{equation}
give, by multiplication on the right,
\begin{gather}
\mathcal D_f(-A+C+P)
=-A^2+C^2-P^2+[A,C]=-\mathcal Q_R,\notag\\
\mathcal Q_R
=\omega^2-p_z^2+\mu_5^2-M^2
-\pi_{f,\perp}^2-iq_fB\gamma^1\gamma^2+2\mu_5H,\qquad H=M\gamma^0\gamma^5-\omega\gamma^5.
\label{eq:app-right-product}
\end{gather}
Accordingly,
$\mathcal D_f^{-1}=(A-C-P)\mathcal Q_R^{-1}$.
The ordering is part of the inverse; in particular, the residual
matrix factor cannot be moved across the transverse Dirac numerator
without accounting for their commutator.

The combination
$\pi_{f,\perp}^2+iq_fB\gamma^1\gamma^2$
has eigenvalue $2n\mathcal B_f$ on $\Pi_n^f$.
Moreover, $H$ commutes with $P_\pm^f$ and
$H^2=\omega^2-M^2$. Defining
$X_n^f=\omega^2-p_z^2+\mu_5^2-M^2-2n\mathcal B_f$,
the second rationalization gives
\begin{align}
U_n^f&=(X_n^f)^2-4\mu_5^2(\omega^2-M^2)=\prod_{a=\pm1}\big[\omega^2-(E_{n,a}^f)^2\big],
\qquad n\geq1,
\label{eq:Xn-def}\\
(X_n^f+2\mu_5H)^{-1}
&=\frac{X_n^f-2\mu_5H}{U_n^f}
\equiv\mathcal G_n^f.
\label{eq:appA-level-resolvent}
\end{align}
For $n=0$ the same polynomial $U_0^f$ can be used as an
intermediate expression; its surplus factor is removed below.

Let $W=A-C$.  Inserting
Eq.~\eqref{eq:appA-landau-projector} into the resolvent yields
\begin{equation}
S_f(\omega,p_z;\bm\rho,\bm\rho')
=i\big[W-\gamma^1\pi_{f,x}-\gamma^2\pi_{f,y}\big]
\sum_{n=0}^{\infty}\Pi_n^f(\bm\rho,\bm\rho')\mathcal G_n^f,
\label{eq:appA-coordinate-resolvent}
\end{equation}
where the differential operators act on the first transverse
coordinate.  Their action on the Schwinger phase is explicit:
\begin{align}
\pi_{f,x}\big[e^{i\Phi_f}F(\bm r)\big]
&=e^{i\Phi_f}\left(-i\partial_{r_x}
-\frac{q_fB}{2}r_y\right)F(\bm r),\notag\\
\pi_{f,y}\big[e^{i\Phi_f}F(\bm r)\big]
&=e^{i\Phi_f}\left(-i\partial_{r_y}
+\frac{q_fB}{2}r_x\right)F(\bm r).
\label{eq:appA-phase-derivatives}
\end{align}
Thus the full coordinate-space propagator factorizes as
$S_f(X,X')=e^{i\Phi_f(\bm\rho,\bm\rho')}
\widetilde S_f(X-X')$.
The constant axial source changes $\mathcal G_n^f$ but does not
alter the Schwinger phase or the oscillator completeness relation.

We Fourier transform the translation-invariant transverse kernel with
$\int d^2r\,e^{-i\bm p_\perp\cdot\bm r}$ and define
$\alpha_f=2p_\perp^2/\mathcal B_f$.
The angular integral gives $2\pi J_0(p_\perp r)$, and the radial identity
\begin{equation}
\int_0^\infty r\,dr\,
e^{-\mathcal B_fr^2/4}
L_m(\mathcal B_fr^2/2)J_0(p_\perp r)
=
\frac{2}{\mathcal B_f}(-1)^m
e^{-p_\perp^2/\mathcal B_f}L_m(\alpha_f)
\label{eq:appA-bessel-laguerre}
\end{equation}
then gives
\begin{equation}
\frac{\mathcal B_f}{2\pi}\int d^2r\,
e^{-i\bm p_\perp\cdot\bm r}e^{-\xi_f/2}L_m(\xi_f)
=2(-1)^m e^{-\alpha_f/2}L_m(\alpha_f).
\label{eq:appA-diagonal-transform}
\end{equation}
Thus the two orbital terms acquire opposite signs after the
replacement $m=n-1$ in the $P_+^f$ sector.
The transverse derivative terms follow from
\begin{equation}
\frac{dL_n(z)}{dz}=-L_{n-1}^{1}(z),\qquad
L_{n-1}(z)+L_{n-2}^{1}(z)=L_{n-1}^{1}(z),
\label{eq:appA-laguerre-derivatives}
\end{equation}
with negative-index polynomials set to zero.  More explicitly, writing
$\bar\pi_{f,x}=-i\partial_{r_x}-q_fBr_y/2$ and
$\bar\pi_{f,y}=-i\partial_{r_y}+q_fBr_x/2$, define
\begin{equation}
F_n^f(\bm r)=e^{-\xi_f/2}
\big[L_n(\xi_f)P_-^f+L_{n-1}(\xi_f)P_+^f\big].
\label{eq:appA-radial-projector}
\end{equation}
The transverse part of the numerator then transforms as
\begin{equation}
-\frac{\mathcal B_f}{2\pi}\int d^2r\,
e^{-i\bm p_\perp\cdot\bm r}
\times(\gamma^1\bar\pi_{f,x}+\gamma^2\bar\pi_{f,y})F_n^f(\bm r)
=
4(-1)^n e^{-\alpha_f/2}L_{n-1}^{1}(\alpha_f)
(\gamma_\perp\!\cdot p_\perp).
\label{eq:appA-transverse-transform}
\end{equation}
Here and below
$\gamma_\perp\!\cdot p_\perp\equiv\gamma^1p_x+\gamma^2p_y$;
this notation does not introduce an additional metric sign.
Equations~\eqref{eq:appA-diagonal-transform} and
\eqref{eq:appA-transverse-transform} account for all three
polynomial structures, $L_n$, $L_{n-1}$, and $L_{n-1}^{1}$.

Together with the ordered resolvent, these transforms give the
spin-sector numerators in Eq.~\eqref{eq:Kn-plus} and
the momentum-space propagator in Eq.~\eqref{eq:propagator-master}.
The ordered factors have the compact form
\begin{equation}
\Lambda_L^f=W(X_n^f-2\mu_5H),\qquad
\Lambda_T^f=(\gamma_\perp\!\cdot p_\perp)(X_n^f-2\mu_5H).
\label{eq:appA-factorized-numerators}
\end{equation}
Expanding these products yields the six longitudinal and three
transverse Dirac structures displayed in
Eqs.~\eqref{eq:LambdaL-corrected} and~\eqref{eq:LambdaT-corrected}.
The terms proportional to $M\omega\gamma^5$ cancel in the product
$WH$, so there is no independent pure-$\gamma^5$ structure.

\subsection{Lowest Landau level and the Feynman pole assignment}
\label{appA:lll-reduction}
For $n=0$, both $L_{-1}$ and $L_{-1}^{1}$ vanish, so
Eq.~\eqref{eq:propagator-master} reduces to
$2ie^{-\alpha_f/2}\Lambda_L^fP_-^f/U_0^f$.
Although the intermediate denominator is quadratic in $\omega^2$,
it cannot represent two physical LLL spin branches.
Indeed,
\begin{equation}
U_0^f=\big[\omega^2-(E_0^f)^2\big]V_0^f,
\qquad
V_0^f=\omega^2-M^2-(p_z-s_f\mu_5)^2,
\label{eq:appA-lll-factorization}
\end{equation}
while the projected numerator satisfies
\begin{align}
\gamma^0\gamma^5P_-^f&=-s_f\gamma^3P_-^f,\qquad
\gamma^3\gamma^5P_-^f=-s_f\gamma^0P_-^f,\qquad
\gamma^0\gamma^3\gamma^5P_-^f=-s_fP_-^f,
\label{eq:appA-lll-identities}\\
\left.\Lambda_L^fP_-^f\right|_{n=0}
&=V_0^f\big[M+\omega\gamma^0
-(p_z+s_f\mu_5)\gamma^3\big]P_-^f.
\label{eq:appA-lll-numerator-factor}
\end{align}
The common factor $V_0^f$ therefore cancels, yielding the reduced LLL
propagator in Eq.~\eqref{eq:LLL-propagator-reduced}, with a single
branch over $p_z\in\mathbb R$.
At $\mu_5=0$, the common factor $X_n^f$ in
Eq.~\eqref{eq:appA-factorized-numerators} cancels one factor of
$U_n^f=(X_n^f)^2$, recovering the usual zero-source Landau propagator
with one LLL spin state and two degenerate spin states for $n\geq1$.
The remaining energy poles carry the Feynman prescription
\begin{equation}
\frac{1}{\omega^2-E^2+i0^+}
\equiv\frac{1}{(\omega-E+i0^+)(\omega+E-i0^+)}.
\label{eq:appA-feynman-prescription}
\end{equation}

\subsection{Coincident trace and the condensate}
\label{app:condensate-reduction}
The coincident scalar trace follows directly from the momentum-space
propagator in Eq.~\eqref{eq:propagator-master} and the orbital density
in Eq.~\eqref{eq:appA-coincident-projector}. The orbital projectors
fix the allowed spin sectors, while integration of the momentum-space
polynomials and the Dirac trace determine the scalar numerator
for each physical energy pole. The transverse measure and polynomial
integral are
\begin{equation}
\frac{d^2p_\perp}{(2\pi)^2}
=\frac{\mathcal B_f}{8\pi}\,d\alpha_f,
\qquad
\int_0^\infty d\alpha\,e^{-\alpha/2}L_m(\alpha)
=2(-1)^m.
\label{eq:appA-transverse-trace-integrals}
\end{equation}
The relevant traces are
\begin{equation}
\operatorname{tr}_D(\Lambda_L^fP_\pm^f)
=2M\big[X_n^f-2\mu_5^2\mp2s_f\mu_5p_z\big],\qquad
\operatorname{tr}_D(\Lambda_T^fP_\pm^f)=0.
\label{eq:appA-numerator-traces}
\end{equation}
For $n\geq1$, integration of the two orbital polynomials adds
the two spin traces, and
\begin{equation}
\frac{4M(X_n^f-2\mu_5^2)}{U_n^f}
=\sum_{a=\pm1}\frac{2M}{\omega^2-(E_{n,a}^f)^2}.
\label{eq:appA-scalar-partial-fractions}
\end{equation}
For $n=0$, only $P_-^f$ survives and its trace contains the
factor $V_0^f$ already removed above.  Hence
\begin{equation}
\int\frac{d^2p_\perp}{(2\pi)^2}
\operatorname{tr}_D\widetilde S_f
={}i\frac{\mathcal B_f}{2\pi}
\left[\frac{2M}{\omega^2-(E_0^f)^2+i0^+}
+\sum_{n=1}^{\infty}\sum_{a=\pm1}
\frac{2M}{\omega^2-(E_{n,a}^f)^2+i0^+}\right].
\label{eq:appA-integrated-scalar-trace}
\end{equation}
There is no additional spin-degeneracy factor multiplying the
explicit branch sum.

After continuation to fermionic Matsubara frequencies,
$\omega_\ell=(2\ell+1)\pi T$, the thermal sum is
\begin{equation}
T\sum_{\ell\in\mathbb Z}\frac{1}{\omega_\ell^2+E^2}
=\frac{1-2n_F(E)}{2E}.
\label{eq:appA-matsubara-sum}
\end{equation}
Including the color multiplicity and the fermionic sign in the
condensate gives
\begin{equation}
\langle\bar\psi_f\psi_f\rangle
={}-N_c\frac{\mathcal B_f}{2\pi}
\int_{-\infty}^{\infty}\frac{dp_z}{2\pi}
\left[\frac{M[1-2n_F(E_0^f)]}{E_0^f}
+\sum_{n=1}^{\infty}\sum_{a=\pm1}
\frac{M[1-2n_F(E_{n,a}^f)]}{E_{n,a}^f}\right].
\label{eq:appA-condensate}
\end{equation}

\section{Vacuum calibration and magnetic finite differences}
\label{app:vacuum-calibration}\label{app:uv-details}
This appendix records the magnetic sum-minus-continuum construction,
vacuum calibration, and regulator-boundary terms relevant to the
finite-$\mu_5$ response.
\subsection{Landau sum minus continuum}
\label{app:magnetic-difference}
For $p\geq0$, define the branch pair
\begin{equation}
E_a(s,p)=\sqrt{M^2+(\sqrt{p^2+s}+a\mu_5)^2},\qquad
g(s,p)=\sum_{a=\pm1}\frac{1-2n_F(E_a(s,p))}{E_a(s,p)}.
\label{eq:magnetic-branch-function}
\end{equation}
Folding the full signed LLL momentum onto $p\geq0$ gives $g(0,p)$;
for $n\geq1$, both branches are already included in $g$ at each
$p_z$, and the same folding gives $2g(2n\mathcal B_f,p)$.
The Landau sum minus the continuum integral is therefore
\begin{equation}
\Delta J_{1,\mathrm{mag}}={}
\frac{N_c}{4\pi^2}\sum_f\int_0^\infty dp\,
\bigg\{\mathcal B_f\bigg[g(0,p)+2\sum_{n=1}^{\infty}g(2n\mathcal B_f,p)\bigg]
-\int_0^\infty ds\,g(s,p)\bigg\}.
\label{eq:MFDS-gap-new}
\end{equation}
With $\delta_f=2\mathcal B_f$, the braces form the trapezoidal
sum $\delta_f[g(0,p)/2+\sum_{n\geq1}g(n\delta_f,p)]$ minus the
continuum integral. The endpoint half weight follows from the
single LLL spin sector, not from an additional degeneracy factor.

The vacuum Landau sum and continuum integral are separately
ultraviolet divergent, so their difference must be defined with a
common subtraction. We multiply $g$ in both expressions by the
auxiliary convergence factor
$g_\epsilon(s,p)=e^{-\epsilon(p^2+s)}g(s,p)$, with $\epsilon>0$.
The variable $s$ is the transverse momentum squared in the continuum
and takes the values $2n\mathcal B_f$ in the Landau sum. Thus both
terms are regulated by the same function of the ultraviolet momentum
before subtraction. With $\mathcal Q_{f,\epsilon}(p)$ denoting the
regulated expression in braces, the prescription is
\[
\Delta J_{1,\mathrm{mag}}=\frac{N_c}{4\pi^2}\sum_f
\lim_{\epsilon\to0^+}\int_0^\infty dp\,\mathcal Q_{f,\epsilon}(p).
\]
This factor is a temporary device for defining the common difference,
not an additional NJL cutoff: its regulator scale is removed after
subtraction. The existence of that limit follows from the ultraviolet
cancellation, rather than from the numerical integration procedure.
At fixed $M>0$, $T$, $\mu_5$, and $\mathcal B_f$, Euler--Maclaurin
summation gives, at sufficiently large $p$, the endpoint-derivative
form of the trapezoidal sum-minus-integral difference.
Since $g(s,p)=2(p^2+s)^{-1/2}+O((p^2+s)^{-3/2})$ in the ultraviolet,
$\partial_sg(0,p)=-p^{-3}+O(p^{-5})$. After removing the auxiliary
factor, the resulting integrand has the asymptotic tail
\begin{equation}
\mathcal Q_f(p)\sim-\frac{\delta_f^2}{12}\,\partial_sg(0,p)
\sim\frac{\mathcal B_f^2}{3p^3}.
\label{eq:detail-magnetic-uv-tail}
\end{equation}
For the chosen Gaussian factor, the additional derivative term
$-\epsilon g(0,p)$ has an integrated ultraviolet contribution of
order $\epsilon|\ln\epsilon|$, which vanishes as $\epsilon\to0^+$.
The longitudinal difference is therefore ultraviolet convergent;
the thermal parts are exponentially suppressed. At nonzero magnetic
fields we evaluate the full difference in
Eq.~\eqref{eq:MFDS-gap-new}, not a weak-magnetic-field expansion.
An $M$-independent electromagnetic normalization of the absolute
potential is treated separately.

\subsection{Vacuum pole and current normalization}
\label{app:detail-vacuum-inputs}
At $T=B=\mu_5=0$, the vacuum calibration uses radial momentum
integrals over $0\leq p<\infty$; the gap equation gives
$\langle\bar uu\rangle=(m_0-M_0)/(4G)$ for either flavor.
With $E_p=\sqrt{p^2+M_0^2}$ and
$E_{p,i}=\sqrt{p^2+M_0^2+a_i\Lambda^2}$, the vacuum kernels are
\[
J_{1,\mathrm{SC}}^{\rm vac}=\frac{N_cN_f}{\pi^2}\int_0^\infty dp\,
\frac{p^2f_\Lambda(p^2)}{E_p},\qquad
J_{1,\mathrm{PV}}^{\rm vac}=\frac{N_cN_f}{\pi^2}\int_0^\infty dp\,p^2
\sum_i\frac{C_i}{E_{p,i}}.
\]
In the random-phase approximation, take the inverse pion denominator
as $1-2G\Pi_{\pi,r}(q^2)$, where $q$ is the pion energy at rest
\cite{Klevansky:1992qe,Buballa:2003qv}. With the same vacuum regulator in
the gap and polarization integrals,
\begin{equation}
\Pi_{\pi,r}(q^2)=J_{1,r}^{\rm vac}+J_{m,r}(q),\qquad
1-2G\Pi_{\pi,r}(q^2)=\frac{m_0}{M_0}-2GJ_{m,r}(q).
\label{eq:detail-pion-gap-identity}
\end{equation}
The finite-energy pion-pole condition is
$0=m_0/M_0-2GJ_{m,r}(m_\pi)$, with $0<m_\pi<2M_0$,
and its kernels are
\begin{align}
J_{m,\mathrm{SC}}(q)&=\frac{N_cN_f}{\pi^2}\int_0^\infty dp\,p^2 f_\Lambda(p^2)
\frac{q^2}{E_p(4E_p^2-q^2)},\notag\\
J_{m,\mathrm{PV}}(q)&=\frac{N_cN_f}{\pi^2}\sum_iC_i\int_0^\infty dp\,p^2
\frac{q^2}{E_{p,i}(4E_{p,i}^2-q^2)}.
\label{eq:vacuum-calibration-pole}
\end{align}
We normalize the axial current
$A_\mu^a=\bar\psi\gamma_\mu\gamma^5\tau^a\psi/2$ through
$\langle0|A_\mu^a|\pi^b(q)\rangle=i\delta^{ab}f_\pi q_\mu$,
using its leading-derivative normalization at zero external momentum:
\begin{equation}
f_{\pi,r}^{\,2}=M_0^2
\left.\frac{\partial\Pi_{\pi,r}(q^2)}{\partial q^2}\right|_{q^2=0}.
\label{eq:detail-fpi-zero-momentum}
\end{equation}
The pion pole is evaluated at finite energy, whereas the stated
$f_\pi$ convention uses the zero-momentum derivative rather than
the finite-energy pole residue. The corresponding calibration
integrals are
\begin{equation}
f_{\pi,\mathrm{SC}}^2=\frac{N_cN_f}{4\pi^2}\int_0^\infty dp\,
\frac{p^2M_0^2 f_\Lambda(p^2)}{E_p^3},\qquad
f_{\pi,\mathrm{PV}}^2=\frac{N_cM_0^2}{4\pi^2}
\ln\frac{(M_0^2+\Lambda^2)^2}{M_0^2(M_0^2+2\Lambda^2)}.
\label{eq:vacuum-calibration-fpi}
\end{equation}
For PV, setting $N_f=2$ gives the expression above;
$\sum_iC_i=0$ removes the common logarithmic endpoint.
The vertex factor remains the physical $M_0^2$ in both
regularizations. For the common-$M_0$ recalibration family of Sec.~\ref{sec:matching-robustness},
$M_0$ replaces the condensate as the third fit target.

\subsection{Finite-\texorpdfstring{$\mu_5$}{mu5} momentum shifts and regulator boundaries}
\label{app:detail-momentum-shifts}
A convergent full-line integral is unchanged by a longitudinal
momentum shift, provided the regularization preserves translation
invariance. The direct SC treatment instead uses a smooth weight
$w(p^2/\Lambda^2)$ evaluated at the unshifted longitudinal momentum
in the LLL. This produces a regulator-gradient contribution to its
vacuum response. Let $C_f=N_c\mathcal B_f/(4\pi^2)$ and
$E_p=\sqrt{M^2+p^2}$. At fixed $M>0$, define
$\delta\Omega_{f,\mathrm{LLL}}^w=-C_f\mu_5^2\mathcal C_E^w
+\mathcal O(\mu_5^4)$ and
$\delta J_{1,f}^{\mathrm{LLL},w}=C_f\mu_5^2\mathcal C_J^w
+\mathcal O(\mu_5^4)$, with $\delta$ denoting the change from
$\mu_5=0$. Integration by parts gives
\begin{align}
\mathcal C_E^w&=-\frac1{\Lambda^2}\int_{-\infty}^\infty dp\,
\frac{p^2}{\sqrt{p^2+M^2}}w'(p^2/\Lambda^2),\notag\\
\mathcal C_J^w&=\frac1{\Lambda^2}\int_{-\infty}^\infty dp\,
\frac{p^2}{(p^2+M^2)^{3/2}}w'(p^2/\Lambda^2)
=\frac1M\partial_M\mathcal C_E^w.
\label{eq:lll-smooth-window}
\end{align}
The prime differentiates the dimensionless argument. For the
SC weight $w(u)=(1+u^5)^{-1}$ of Eq.~\eqref{eq:soft-window-new},
$w'(u)=-5u^4/(1+u^5)^2\leq0$; hence
$\mathcal C_E^w>0$ and $\mathcal C_J^w<0$. Unlike an actual finite
integration endpoint, the smooth-weight contribution arises where
$w'$ is nonzero. The familiar hard-window coefficients follow from
the formal limit $w(u)=\Theta(1-u)$:
$\mathcal C_E^w=\Lambda/E_\Lambda$ and
$\mathcal C_J^w=-\Lambda/E_\Lambda^3$, with
$E_\Lambda=\sqrt{M^2+\Lambda^2}$; no hard window is used in the
present SC calculations.

For comparison, the direct PV prescription first combines regulator
masses $M_i^2=M^2+a_i\Lambda^2$ with coefficients $C_i$ as specified
in Eq.~\eqref{eq:PV-data-new}. The complete PV-regulated LLL vacuum
potential at fixed $M$ is proportional to the full-line integral
\begin{equation}
\Omega_{f,\mathrm{LLL}}^{\mathrm{PV}_{\mathrm D}}(\mu_5)
=-C_f\int_{-\infty}^{\infty}dp\,
\sum_iC_i\sqrt{M_i^2+(p+s_f\mu_5)^2}
=\Omega_{f,\mathrm{LLL}}^{\mathrm{PV}_{\mathrm D}}(0).
\label{eq:lll-pv-shift-invariance}
\end{equation}
Here $\sum_i C_i=0$ and $\sum_iC_iM_i^2=0$ cancel the leading
ultraviolet terms of the combined integrand, leaving an
$O(|p|^{-3})$ tail. The convergent PV combination may therefore be
shifted by $p\mapsto p+s_f\mu_5$ without generating an endpoint
contribution. Its corresponding LLL vacuum gap kernel is also
independent of $\mu_5$ at fixed $M$. This is a property of the
complete direct PV-regulated LLL vacuum integral, not of its
unregularized mass terms taken separately.

The SC--PV distinction here concerns only this LLL vacuum
contribution. It does not determine the finite-$\mu_5$ dependence
of the higher Landau levels or the full equilibrium gap equation.
In MFDS, the common magnetic contribution is defined by the entire
Landau sum-minus-continuum difference in
Eq.~\eqref{eq:MFDS-gap-new}; the isolated direct-PV LLL argument
must not be applied to that difference term by term.

\section{Finite-\texorpdfstring{$\mu_5$}{mu5} matching, scalar thermodynamics, and crossover responses}
\label{app:source-matching-details}
This appendix documents the independent matching input, the scalar
susceptibility used for criterion (2), and the local derivative
relations for criterion (1).

\subsection{Finite-\texorpdfstring{$\mu_5$}{mu5} matching input and model spread}
\label{app:finite-source-input}
The adopted $\mathcal R_\chi(0)=1.0057$ is digitized at
$\mu_5=150\,\mathrm{MeV}$ from the green triangles in the left panel of
Fig.~3 of Ref.~\cite{GomezNicola:2023ghi}; $[1.0035,1.0079]$
is a nonstatistical sensitivity range. We select the green series because
it uses the scalar-susceptibility peak, matching criterion (2) here.
The blue squares instead use the NNLO ChPT condensate-vanishing
condition~\cite{Espriu:2020dge}, whereas the red circles are
$N_c=2$, $N_f=4$ lattice results with heavier-than-physical
pions~\cite{Braguta:2015zta}.
The two theory works share the same lattice-constrained fit, so their
proximity to the red points is not an independent check.
The unitarized ChPT calculation uses
$m_\pi=140\,\mathrm{MeV}$ and $f_\pi=93\,\mathrm{MeV}$, versus
$138\,\mathrm{MeV}$ and $92.4\,\mathrm{MeV}$ in the present
vacuum calibration. We use the equal-magnitude source convention
specified in Sec.~\ref{sec:source-input}.

For a common input $y$, the baseline coefficients satisfy
\begin{equation}
\mathcal R_\chi(0;c_r(y))=y,\qquad r=\mathrm{SC,PV}.
\label{eq:detail-common-input-match}
\end{equation}
Because D and MFDS coincide at zero magnetic field ($B=0$), each
vacuum regularization has one coefficient shared by its two magnetic-field
treatments. At finite magnetic fields, write
$\mathcal R_\chi^{(s)}(eB;y)=\mathcal R_\chi^{(s)}(eB;c_{r(s)}(y))$.
The spread and its contraction factor are
\begin{equation}
S(eB;y)
=\max_s\mathcal R_\chi^{(s)}(eB;y)
-\min_s\mathcal R_\chi^{(s)}(eB;y),\quad
\eta(eB;y)=\frac{S_{\rm before}(eB)}
{S(eB;y)}.
\label{eq:detail-common-input-spread}
\end{equation}
Here $S_{\rm before}$ is the unmatched four-treatment spread,
and the contraction factor is defined when the denominator is nonzero.
The three common inputs are correlated sensitivity scenarios, not
statistical confidence limits or bounds over a continuous input range.

\begin{table*}[!htbp]
\caption{Matched spread $S$ and its reduction
$\eta=S_{\rm before}/S$ for three choices of the common
matching input at zero magnetic field ($B=0$). The $eB$ values are in $\mathrm{GeV}^2$. Each spread uses all
four treatments at the same input. The alternative inputs are
nonstatistical sensitivity scenarios, not a continuous uncertainty bound.}
\label{tab:polish-spreads}
\begin{ruledtabular}
\begin{tabular}{c cc cc cc}
 & \multicolumn{2}{c}{$eB=0.1$} & \multicolumn{2}{c}{$eB=0.2$} & \multicolumn{2}{c}{$eB=0.4$}\\
$y$ & $10^3 S$ & $\eta$ & $10^3 S$ & $\eta$ & $10^3 S$ & $\eta$\\
\hline
1.0035 & 2.119 & 74.91 & 5.383 & 26.85 & 8.303 & 13.70\\
1.0057 & 2.138 & 74.24 & 5.440 & 26.58 & 8.352 & 13.62\\
1.0079 & 2.157 & 73.59 & 5.496 & 26.30 & 8.400 & 13.54\\
\end{tabular}
\end{ruledtabular}
\end{table*}

At the three sampled magnetic fields (Table~\ref{tab:polish-spreads}),
the matched spread differs
from its central-input value by at most about $1.04\%$ across these
discrete common-input scenarios. Substantial spread reduction relative
to $c_r=0$ therefore persists for the tested inputs.

\subsection{Scalar susceptibility from the matched potential}
\label{app:detail-equilibrium-thermodynamics}\label{app:matching-operators}
For a smooth stable equilibrium $\Omega_M(M_*;y,z)=0$, the envelope
identities for external parameters $y,z$ are
\begin{equation}
\frac{d\Omega_*}{dy}=\Omega_y,\qquad
\frac{d^2\Omega_*}{dy\,dz}
=\Omega_{yz}-\frac{\Omega_{yM}\Omega_{Mz}}{\Omega_{MM}},
\qquad \Omega_{MM}>0.
\label{eq:envelope-new}
\end{equation}
The partial derivatives on the right are evaluated at $M=M_*$.
An $M$-independent normalization can affect absolute source
derivatives without changing the equilibrium mass.

For a fixed treatment, we suppress its treatment superscript in this
subsection. For the matching observable,
the potential convention is
\begin{align}
\Omega^{(c)}(M;T,B,x,m_0)
&=\frac A2(M-m_0)^2
+\Omega_q(M;T,B,x)+\mathcal N(T,B,x),\notag\\
A&=\frac1{2G}+2c_rx,\qquad x=\mu_5^2.
\label{eq:detail-matched-potential}
\end{align}
Here $\partial_M\Omega_q=-MJ_1$, and $G,\Lambda,c_r$ are fixed
under $m_0$ differentiation. We take $\mathcal N$ to be
independent of $m_0$, so $\Omega_{Mm_0}^{(c)}=-A$ and
$\Omega_{m_0m_0}^{(c)}=A$. Thus
\[
\chi_m^{(c)}=\frac{\partial M_*}{\partial m_0}
=\frac{A}{R_M^{(c)}},\qquad
\chi_{\rm scal}^{(c)}=-\frac{d^2\Omega_*^{(c)}}{dm_0^2}
=\frac{A^2}{R_M^{(c)}}-A
=A\bigl(\chi_m^{(c)}-1\bigr).
\]
The numerical scalar susceptibility uses this expression, including
the $-A$ term. Relative potential differences are used only for
equilibrium-branch selection, not for current-mass differentiation.
At fixed $x$, $A$ is temperature independent, and $A>0$ ensures
that the two susceptibilities share a peak location. If an additional
$M$-independent term depends on $m_0$, its second mass derivative
would enter $\chi_{\rm scal}$; Eq.~\eqref{eq:detail-matched-potential}
fixes the convention used here.

\subsection{Implicit derivatives and crossover criteria}
\label{app:detail-local-derivatives}\label{app:crossover-response}

\begin{figure*}[!htbp]
\centering
\includegraphics[width=.97\textwidth]{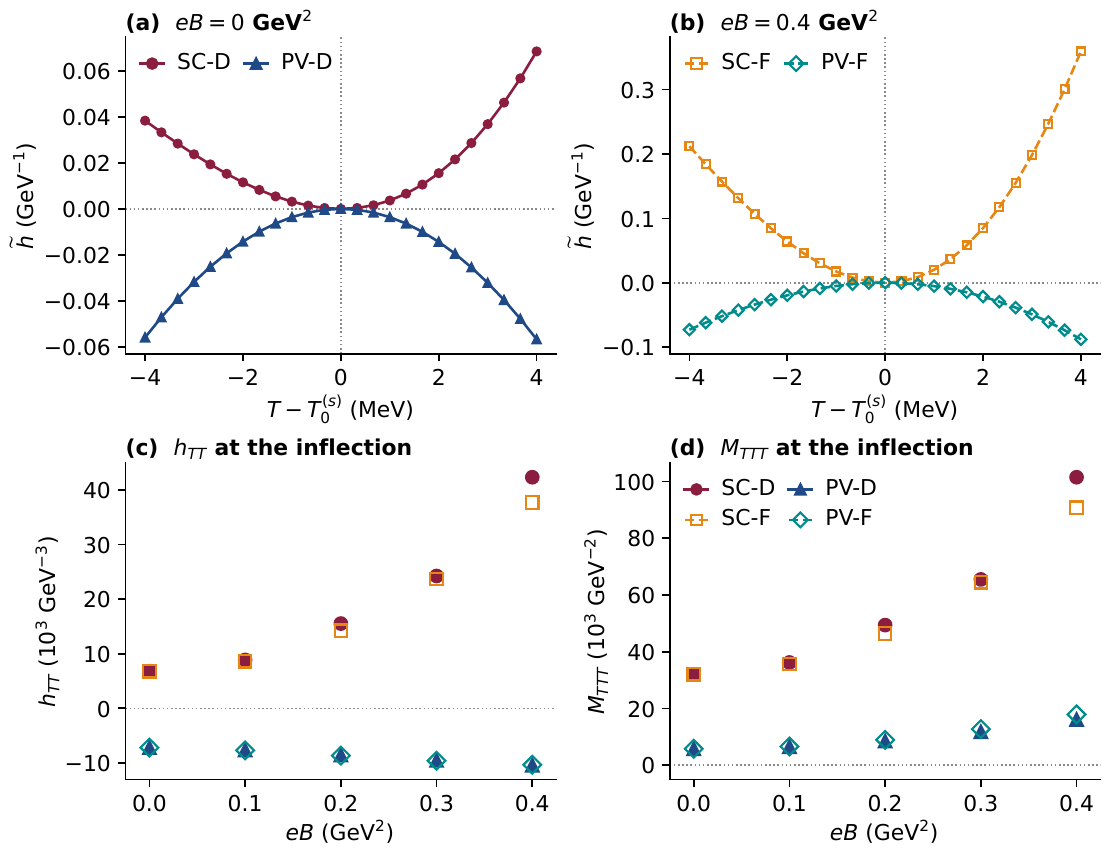}
\caption{Temperature curvature of the equilibrium-mass response $h(T)=\left.\partial M_*/\partial\mu_5^2\right|_{\mu_5=0}=-R_x/R_M$ on the unmatched scalar branch. Panels (a,b) show $\widetilde h(T)=h(T)-h(T_0)-h_T(T_0)(T-T_0)$ around each model's $\mu_5=0$ mass inflection $T_0$, with 25 temperatures per curve over $\pm4\,\mathrm{MeV}$ and separate vertical scales. Panels (c,d) give $h_{TT}$ and $M_{TTT}$ at the 20 reference-calibration anchors, covering $eB=0,0.1,0.2,0.3,0.4\,\mathrm{GeV}^2$ in each treatment. D and F coincide at zero field. At these anchors $M_{TTT}>0$, so the local shift $K_5=-h_{TT}/M_{TTT}$ is controlled by the temperature curvature of $h$.}
\label{fig:polish-local-curvature}
\end{figure*}

At fixed magnetic field and model parameters, including any source coefficient,
the equilibrium mass obeys $R(M_*(T,x),T,x)=0$, $x=\mu_5^2$,
where $R$ is the original or extended residual, as appropriate.
Partial derivatives of $R$ are evaluated at independent $M,T,x$ and
then on the selected smooth stable branch, $R_M=\Omega_{MM}>0$.
Set $v=M_T$, $h=M_x$, $u=M_{TT}$, and $w=M_{Tx}$, with the
total-derivative operators
\begin{equation}
\mathcal D_T=\partial_T+v\partial_M,\qquad
\mathcal D_x=\partial_x+h\partial_M.
\label{eq:detail-total-derivatives}
\end{equation}
Implicit differentiation gives
\begin{align}
v&=-\frac{R_T}{R_M},\qquad h=-\frac{R_x}{R_M},\notag\\
R_Mu&=-R_{TT}-2R_{MT}v-R_{MM}v^2,\notag\\
R_Mw&=-R_{Tx}-R_{MT}h-R_{Mx}v-R_{MM}vh.
\label{eq:implicit-low-order}
\end{align}
The required third derivatives are obtained without a separate
expanded chain rule:
\begin{equation}
M_{TTT}=\mathcal D_Tu,\qquad M_{TTx}=\mathcal D_xu.
\label{eq:third-implicit-chain}
\end{equation}
Here $u$ denotes the right-hand side of its relation in
Eq.~\eqref{eq:implicit-low-order} divided by $R_M$, with
$v=-R_T/R_M$ substituted. The operators act on this full function
of independent $M,T,x$, including the denominator, before evaluation
on $M=M_*(T,x)$. The inflection condition $u=0$ is imposed only
after differentiation; $\mathcal D_x$ holds $T$ fixed.

For criterion (1), the selected nondegenerate interior maximum of
$|M_T|$ on a decreasing branch has $u=0$, $v<0$,
and $M_{TTT}>0$. Differentiating $u(T_{\rm pc}^{(1)}(x),x)=0$
along the crossover curve gives $0=M_{TTT}K_5+M_{TTx}$, hence
$K_5=-M_{TTx}/M_{TTT}$, or $K_5=-h_{TT}/M_{TTT}$ at $x=0$.
For the unmatched reference calibrations, Fig.~\ref{fig:polish-local-curvature}
illustrates how the temperature curvature $h_{TT}$ controls the sign of
this local shift when $M_{TTT}>0$.

For the independently matched model, the additional interaction
vanishes at $x=0$. The original and matched equilibrium trajectories
therefore share the same $M_*(T,0)$, $R_{0,M}$, and $M_{TTT}$.
Combining the local relation above with
Eq.~\eqref{eq:matched-mass-response} gives
\begin{equation}
K_5^{(c)}(eB,0)-K_5^{(0)}(eB,0)
=\left.\frac{2c_r}{M_{TTT}}\frac{d^2}{dT^2}
\left(\frac{M_*(T,0)-m_0}{R_{0,M}}\right)
\right|_{T=T_{\rm pc}^{(1)}(eB,0)}.
\label{eq:matched-local-K5-shift}
\end{equation}
The temperature derivatives act on the complete ratio along the
common zero-$\mu_5$ equilibrium trajectory.

For criterion (2), the derivative along the equilibrium mass
trajectory is $\mathcal D_TR_M=R_{MT}+vR_{MM}$, with $v=M_T$.
The temperature independence of $A=(2G)^{-1}+2c_rx$ at fixed $x$
then gives the susceptibility peak conditions, when
$A>0$ and $R_M^{(c)}>0$,
\begin{equation}
\mathcal D_TR_M^{(c)}=0,\qquad
\mathcal D_T^2R_M^{(c)}>0.
\label{eq:detail-susceptibility-locator}
\end{equation}
This is the analytic condition for the common susceptibility peak
under the convention of Appendix~\ref{app:matching-operators}.
Numerically, we maximize $\chi_m$ for the original-baseline criterion
comparison and $\chi_{\rm scal}^{(c)}$ for the matching and peak ratios.
All these formulas require
an interior nondegenerate extremum on a smooth branch; crossings and
endpoint maxima must be evaluated by their defining conditions.

\FloatBarrier
%
\makeatother
\end{document}